\documentclass[journal,twoside,web]{ieeecolor}
\usepackage{tmi}
\usepackage{cite}
\usepackage{amsmath,amssymb,amsfonts}
\usepackage{amsmath}
\usepackage{soul}
\usepackage{algorithm}
\usepackage[noend]{algpseudocode}
\usepackage{caption}
\usepackage{makecell}
\usepackage{todonotes}

\DeclareMathOperator*{\argmin}{arg\,min}

\usepackage{hyperref}
\hypersetup{
    colorlinks=true,
    linkcolor=blue,
    filecolor=magenta,      
    urlcolor=black,
    pdftitle={Overleaf Example},
    pdfpagemode=FullScreen,
    }

\DeclareMathAlphabet{\pazocal}{OMS}{zplm}{m}{n} 

\usepackage{graphicx}
\usepackage{textcomp}
\usepackage[draft]{changes}
\definechangesauthor[name={J. Martin}, color=blue]{JBM}

\def\BibTeX{{\rm B\kern-.05em{\sc i\kern-.025em b}\kern-.08em
    T\kern-.1667em\lower.7ex\hbox{E}\kern-.125emX}}
\begin{document}

This work has been submitted to the IEEE for possible publication. Copyright may be transferred without notice, after which this version may no longer be accessible
\clearpage

\title{ {A Minibatch Alternating Projections Algorithm for Robust and Efficient Magnitude Least-Squares RF Pulse Design in MRI}}
\author{Jonathan B. Martin, Charlotte R. Sappo \IEEEmembership{Student Member, IEEE}, Benjamin M. Hardy, and William A. Grissom, \IEEEmembership{Member, IEEE}
\thanks{Manuscript received XX; revised XX; accepted
XX. Date of publication XX; date of
current version XX. This work was supported in part by the U.S. NIH under Grant R01 EB016695.}
\thanks{J. B. Martin is with the Department of Biomedical Engineering, Vanderbilt University, Nashville, TN 37205 USA (e-mail:jonathan.b.martin@vanderbilt.edu). }
\thanks{C. R. Sappo is with the Department of Biomedical Engineering, Vanderbilt University, Nashville, TN 37205 USA (e-mail:charlotte.r.sappo@vanderbilt.edu). }
\thanks{B. M. Hardy is with Remcom, Inc., State College, PA 16801 USA (email:benjamin.hardy@remcom.com).}
\thanks{W. A. Grissom is with the Department of Biomedical Engineering,
Case Western Reserve University, Cleveland, OH 44106 USA (e-mail: wag57@case.edu).}}

\maketitle
\begin{abstract}
A magnitude-least-squares radiofrequency pulse design algorithm is reported which uses interleaved exact and stochastically-generated inexact updates to escape local minima and find low-cost solutions. Inexact updates are performed using a small randomly selected minibatch of the available $B_1^+$ measurements to update RF pulse weights, 
which perturbs the sequence of alternating projections.
Applications to RF shimming, parallel transmit spokes RF pulse design,  {and spectral-spatial RF pulse design} are considered. 
Numerical and simulation studies characterized the optimal minibatch size, 
which was found to consistently produce lower power and lower RMSE solutions across subjects, coil geometries, $B_1^+$ resolutions and orientations. 
The method was validated in-vivo at 7 Tesla and produced improvements in image quality in a slice-by-slice RF-shimmed 
imaging sequence. 
Compared to conventional methods, 
the pulse design method can more robustly design RF pulses that correct for $B_1^+$ inhomogeneities at ultra-high field strengths, 
and enable pulse designs to be completed with increased computational efficiency.
\end{abstract}

\begin{IEEEkeywords}
RF pulse design, RF shimming, phase retrieval, high-field magnetic resonance imaging (MRI), parallel transmission (pTx), magnitude least squares, Gerchberg-Saxton, alternating projections, minibatch
\end{IEEEkeywords}

\section{Introduction}
\label{sec:introduction}
 { \IEEEPARstart{I}{n} magnetic resonance imaging (MRI), as the strength of the main magnetic field increases, 
the wavelength of the RF transmit field decreases. 
This can result in transmit field inhomogeneities and standing waves in higher field strength scanners
as the RF wavelength approaches the size of the imaged object \cite{Yang2002AnalysisField, Collins2003SpatialCoil, Alecci2001RadioProperties}. 
These transmit field inhomogeneities produce large variations in image intensity and tissue contrast, 
which have a detrimental effect on image quality \cite{Sengupta2017Evaluation3T}. 
A widely-used approach is to correct or compensate inhomogeneities by design of the driving amplitudes and phases of an MRI scanner's multiple transmit coils.
This may be achieved with either static RF shims, 
in which the transmit field's driving amplitude and phase are designed but stay fixed during transmit \cite{Ibrahim2001Effects, Adriany2005TransmitImaging, Mao2006ExploringHead}, 
or with dynamic RF pulses, 
in which driving amplitude and phase vary temporally, 
often in coordination with dynamically varying magnetic field gradients \cite{Katscher2002, Zhu2002, Setsompop2008spokes, Malik2012TailoredPulses, Martin2020, Williams2023Ultra-highDesign}. 
The optimization of either RF shims or pulses for these methods is often collectively referred to as parallel transmission (pTx) RF pulse design, 
and either technique can make signal intensity more uniform and improve image quality. }

\par When designing RF shims or pulses to mitigate transmit field inhomogeneity, 
phase variations across the excited volume are permissible when they do not lead to signal loss \cite{Schricker2001DualbandImaging, Mao2006ExploringHead, Setsompop2008MagnitudeChannels}. 
By allowing the excitation phase profile to freely vary
when designing RF shims or pulses, 
greatly improved homogeneity of the magnitude of the excitation profile can be achieved \cite{Setsompop2008MagnitudeChannels}. 
This design preference can be captured by formulating the pulse design problem as a magnitude-least-squares (MLS) optimization problem \cite{Kassakian2006ConvexSynthesis, Setsompop2008MagnitudeChannels}. 
 { The MLS framing of RF pulse design is equivalent to a phase retrieval problem \cite{Fienup1982PhaseComparison, Wong1995TheProblem}, 
which is commonly encountered in other fields of imaging \cite{Yan2013AApplications, Li2021Patch-BasedImaging, Shechtman2015PhaseOverview}.
The MLS RF pulse design problem in MRI is the recovery of an unknown complex-valued RF pulse and the magnetization phase corresponding to a uniform excitation flip angle magnitude across the region to be imaged \cite{Hoyos-Idrobo2014OnConstraints}.}

\par The phase retrieval MLS RF pulse design problem is commonly solved using variants of the Gerchberg-Saxton phase retrieval algorithm \cite{Gerchberg1972}, also known as the variable-exchange method \cite{Setsompop2008MagnitudeChannels, Kassakian2006ConvexSynthesis, Hoyos-Idrobo2014OnConstraints}.
It is nearly ubiquitously used for multidimensional pulse designs \cite{Paez2021RobustRegularization, Cloos2012KT-points:Volume, Herrler2021FastOptimization}, 
or to provide initialization for constrained designs \cite{Hoyos-Idrobo2014OnConstraints, Gras2017UniversalTransmission}. 
Gerchberg-Saxton is favored for these applications due to its simplicity and reduced computational complexity versus semidefinite program relaxations \cite{ Candes2013PhaseCompletion} or interior-point methods \cite{ Helmberg1996AnProgramming}. 
However, as a special case of the method of alternating projections \cite{Noll2015OnProjections}, 
the Gerchberg-Saxton method only guarantees local convergence \cite{Noll2021AlternatingReduction}, 
and thus is prone to converging on suboptimal pulse solutions, 
some of which may produce excitation with severe inhomogeneities or even complete signal nulls \cite{Paez2021RobustRegularization}. 
Regularization terms may be added to the optimization's cost function to penalize null solutions, 
but this may increase excitation error \cite{Paez2021RobustRegularization}. 
Oftentimes, pulse designers resort to multiple trials of the algorithm with different random initializers in hopes that one
initializer will lead to a global minimum \cite{Hoyos-Idrobo2014OnConstraints, Wu2018High-resolutionTransmission}. 
However, repeating the optimization with many initial points
is computationally inefficient, especially if substantial progress has already been made towards feasible points in previous trials.

\par As an alternative, 
we investigate a modified version of Gerchberg-Saxton MLS pulse design algorithm which avoids the premature convergence to high-cost local minima of the conventional Gerchberg-Saxton approach by introducing stochasticity into the algorithm. 
This stochasticity is achieved by performing MLS optimization steps with a small random minibatch of the available $B_1^+$ data.  {Minibatch sampling has been previously explored in the setting of alternating projections updates \cite{Necora2019}, 
where it has (for example) been shown to enable parallelization strategies \cite{Asi2020}. 
Minibatching has also been analyzed in the settings of nonconvex optimization \cite{Ghadimi2014} and  phase retrieval problems \cite{Davis2019} similar to the MLS pulse design problem, 
where it was shown to provide complexity and convergence guarantees. Introducing noise into updates via minibatching or stochastic gradient descent can also help avoid convergence to high-cost local minima \cite{KleinbergSGD2018, Goodfellow2016}. 
We intend to translate these previously demonstrated benefits of minibatching to MLS RF pulse design in MRI.} 
Unlike other phase retrieval applications, 
there is a substantial amount of redundant information in the $B_1^+$ measurements used for MLS RF pulse design due to the fact that $B_1^+$ maps are sampled at a much finer resolution than needed to characterize electromagnetic spatial variation. 
As a result, optimizing over a minibatch also provides a computational acceleration by eliminating redundant measurements. 
The proposed approach offers a more efficient method of exploring feasible points than successive random initializations.  

\par We explore the effect that minibatching row-sampling has on the system, 
demonstrate the algorithm's performance in simulation, 
and validate it with an in-vivo 7T RF shimming experiment. 
Preliminary results supporting this work were reported in Ref. \cite{Martin2023}

\section{Theory}
\subsection{Magnitude-least-squares RF pulse design}
The design of a length-$N_t$, $N_c$ transmit-channel RF pulse for flip angle homogenization of a spatial region with $N_s$ voxels can be expressed as a magnitude-least-squares optimization problem \cite{Hoyos-Idrobo2014OnConstraints}: 

\begin{equation} 
\argmin_{\textbf{x}} \left \Vert \lvert\textbf{A}\textbf{x}\rvert - \textbf{y}\right \Vert ^2_2 +  {\beta} \left \Vert \textbf{x} \right\Vert^2_2, \\ 
 \label{mls}
 \end{equation}

\noindent where  $\textbf{A} \in \mathbb{C}^{N_s \times N_cN_t}$ is the small-tip angle parallel Fourier transmit design matrix \cite{Grissom2006SpatialExcitation}
(or, a matrix of complex-valued multi-channel $B_1^+$ field maps in the related $N_t = 1$ RF shimming case), $\textbf{x}\in \mathbb{C}^{N_cN_t}$ contains the concatenated multichannel RF pulses, 
and $\textbf{y}\in \mathbb{C}^{N_s}$ is the desired target magnetization profile (or, the target shimmed $B_1^+$ field), 
and the $| \cdot |$ operator returns the element-wise absolute value of the scalar elements of a vector. 
 { $\beta$} is an optional regularization parameter penalizing pulse power; other regularization schemes are possible \cite{Hoyos-Idrobo2014OnConstraints, Paez2021RobustRegularization}. 
The representative MLS pulse design problems considered in this work are the one- or multi-spoke multislice RF pulse design problem \cite{Saekho2006Fast-kzInhomogeneity, Setsompop2006ParallelTesla},  {and the multichannel spectral-spatial RF pulse design problem} \cite{MalikWater2010}. 

\par In the Gerchberg-Saxton phase retrieval approach, 
the optimization is reformulated to avoid the inclusion of the absolute value function (which is not differentiable when its argument equals zero) 
by splitting the amplitude and phase of the target magnetization to
$\textbf{y} = \textbf{b} \cdot \textbf{z}$ where $\textbf{b} = |\textbf{y}|$ and $\textbf{z} = e^{\imath \angle\textbf{y}}$ \cite{Kassakian2006ConvexSynthesis, Hoyos-Idrobo2014OnConstraints}. 
The reformulated problem, neglecting regularization, is: 

\begin{align} 
\argmin_{\textbf{x, z}} &||\hspace{0.1cm} \textbf{A}\textbf{x} - \textbf{b}\cdot \textbf{z}||^2_2 \label{reformulation} \\
\textbf{z} \in& \mathbb{C}^{N_s \times 1}, \nonumber \\
|z_i| = 1 \hspace{3mm} &\forall \hspace{1mm} i \in [1 .. N_s] \nonumber
\end{align}
 where $\textbf{b} \cdot \textbf{z}$ is the element-wise product of the two vectors, and $i$ indexes the row dimension. 
\noindent Proof of the equivalence of Eqn. \ref{mls} and Eqn. \ref{reformulation} was provided in \cite{Kassakian2006ConvexSynthesis}.

\par The solution of the problem in Eqn. \ref{reformulation} is an RF pulse/excitation phase pair $(\textbf{x}^*, 
\textbf{z}^*)$, that best satisfies two conditions encapsulated in the optimization problem: Condition 1) that $\textbf{x}$ is the solution of the linear least squares optimization problem; Condition 2) that the vector $\textbf{z}$ is in the set $\mathcal{B}$ with elements of phase equal to that of $\textbf{A}\textbf{x}$ and absolute value 1 for all $i$. \cite{Kassakian2006ConvexSynthesis}.
Progress is made towards a feasible point $(\textbf{x}^*, \textbf{z}^*)$ by a sequence of alternating projections that satisfy each of these conditions in turn. 
After some initializer phase $\textbf{z}$ is determined, 
Condition 1) is met by projecting the current $\textbf{y} = \textbf{b} \cdot \textbf{z}$ on the image of $\textbf{A}$ using the orthogonal projector $\textbf{AA}^\dagger$,
where $\textbf{A}^\dagger$ is the (typically regularized or singular value-truncated) pseudoinverse of $\textbf{A}$. 
This projection is practically computed as the solution to the linear least squares problem of Eqn. \ref{reformulation}:

\begin{equation} \textbf{x}^+ \in P_\mathcal{C}(\textbf{y}) = \textbf{A}^\dagger (\textbf{b} \cdot \textbf{z}).
 \label{projection1}\end{equation}

\noindent The new estimate $\textbf{x}^+$ is then used to update the phase variable of $\textbf{y} =\textbf{b} \cdot \textbf{z}$ to meet Condition 2:
\begin{equation} \textbf{z}^+ = \textbf{Ax}^+/|\textbf{Ax}^+|.
 \label{projection2}\end{equation}
Following the convention used in \cite{Noll2015OnProjections} that within this update that $0/|0| = 1$, 
this mapping is known to be an orthogonal projection $P_\mathcal{B}(\textbf{x})$ on the magnitude set $\mathcal{B}$. \cite{Noll2015OnProjections, Combettes2002PhaseOptimization}. 
This set was shown to be subanalytic and prox-regular \cite{Noll2015OnProjections}.
Gerchberg-Saxton MLS RF pulse design is thus a special case of alternating projections: 
\begin{gather*}\textbf{x}^+ \in P_\mathcal{C}(P_\mathcal{B}(\textbf{x})), \textbf{z}^+ \in P_\mathcal{B}(P_\mathcal{C}(\textbf{b} \cdot \textbf{z}))
\end{gather*}
which generates a sequence $\textbf{x}, \textbf{z}, \textbf{x}^+, \textbf{z}^+, \textbf{x}^{++}, \textbf{z}^{++},...$ until some convergence threshold is reached with a solution pair $(\textbf{x}^*, \textbf{z}^*)$. The full procedure is encapsulated in Algorithm \ref{alg:GSalg}.

\begin{algorithm}
    \captionof{algorithm}[Gerchberg-Saxton MLS Pulse Design (PseudoCode)]{Gerchberg-Saxton RF Pulse Design}
    \begin{algorithmic}[1]
        \State \textbf{Input:} $\textbf{A} \in \mathbb{C}^{N_s \times N_cN_t}$, $\textbf{x}_0\in \mathbb{C}^{N_cN_t}$, $\textbf{b}\in \mathbb{C}^{N_s}$

        \State $\textbf{x} \gets \textbf{x}_0$
        \State $\textbf{z} \gets e^{\imath \angle (\textbf{Ax}_0)}$
        \While{\text{not done}}
            \State $\textbf{y} \gets \textbf{b} \cdot \textbf{z} $
            \State $\textbf{x} \gets (\textbf{A}^H\textbf{A})^{\dagger}\textbf{A}^H\textbf{y}$
            \State $\textbf{z} \gets  (\textbf{Ax})/|\textbf{Ax}|$

            \State \text{done} = $\text{(iter $>$ maxiter)} \parallel \text{(residual $<$ tolerance)}$
            
        \EndWhile
        \State $\text{return} (\textbf{x}^*,\textbf{z}^*)$
    \end{algorithmic}
    \label{alg:GSalg}
\end{algorithm}

\par This approach is locally convergent: 
if $\textbf{z}^*$ is an approximate solution to the phase retrieval problem, 
there exists $\delta > 0$ such that whenever this Gerchberg-Saxton projection sequence enters the closed Euclidean ball around a stationary point $\textbf{z}^*$ it converges towards that stationary point with sublinear rate of convergence \cite{Noll2015OnProjections}.
As a result, this method is unable to escape from local minima after entering their vicinity. The performance of the conventional MLS method for designing RF pulses is thus highly sensitive to its phase initialization.

\subsection{Inexact Alternating Projection}
Our proposed approach to escape early convergence to high-cost local minima in MLS RF pulse design is to insert inexact projections \cite{Lewis2009LocalProjections, Kruger2016RegularityProjections} into the sequence of Gerchberg-Saxton alternating projections. 
An inexact projection is one in which there is some approximation made of the defined exact projection.  
In the context of Gerchbeg-Saxton RF pulse design this could be produced by inexactly solving the linear least squares problem which determines the pulse weights (Eqn. \ref{projection1}) or inaccurately updating the phase variable (Eqn. \ref{projection2}). 

\par Under the condition that these deviations in direction and magnitude of projection are sufficiently small, 
local linear convergence has repeatedly been shown to hold \cite{Lewis2009LocalProjections, Kruger2016RegularityProjections}. However, if the deviations in projection are large enough, 
guarantees of error reduction and local convergence may be violated in inexact iterations.  
We propose to use periodic inexact projections interleaved in the Gerchbeg-Saxton projection sequence to escape local convergence regions and efficiently explore the feasible points of the RF pulse design phase retrieval problem.


\subsection{Random Minibatches of the Spatial Dimension Data to Create Inexact Projections}

In this work, an inexact projection in the Gerchberg-Saxton alternating projection sequence is generated by performing the projection of Equation \ref{projection1} using only a small subset of the rows of $\textbf{A}$ and $\textbf{bz}$. 
This row dimension is the spatial dimension of the pTx system; this corresponds to performing iteration updates using only a small ``minibatch'' of the available $B_1^+$ measurements scattered across space. 
The sampling process can be modeled as the premultiplication of the pTx system matrix $\textbf{A}$ and the target $\textbf{bz}$ by a row-sampling matrix $\textbf{S}$ \cite{Drineas2016RandNLA} prior to performing the projection, $\mathbf{\tilde{A}} = \mathbf{SA}$. 
 For fixed $\textbf{z}$, this generates a perturbed projection 
\begin{equation} \Tilde{\textbf{x}}^+ \in P_\mathcal{C}(\textbf{y}) = \mathbf{\tilde{A}}^\dagger \textbf{S}(\textbf{b} \cdot \textbf{z})
 \label{minibatched_projection}\end{equation}
Equation \ref{minibatched_projection} is the solution to the minibatched least squares approximation problem (for fixed $\textbf{z}$)
\begin{gather} \argmin_{\textbf{x}} \left \Vert \hspace{0.1cm} \textbf{S}(\textbf{A}\textbf{x} - \textbf{b}\textbf{z})\right\Vert^2_2.
 \label{minibatched_optimization}
\end{gather}
\noindent $\textbf{S}$ may sample the row dimension with uniform probability, 
or with some weighted probability importance sampling \cite{Drineas2016RandNLA}; 
in this work we consider uniform-probability row sampling without replacement \cite{Gross2010NoteMatrices}. 



Taking the inexact projection perturbs the alternating projection sequence by an amount
\begin{gather} ||\textbf{x}^+-\tilde{\textbf{x}}^+|| = ||\textbf{A}^\dagger (\textbf{b} \cdot \textbf{z}) - \mathbf{\tilde{A}}^\dagger \textbf{S}(\textbf{b} \cdot \textbf{z})||
\end{gather}
If $\textbf{S}$ fully samples the row dimension, then $||\textbf{x}^+-\tilde{\textbf{x}}^+|| = 0$, producing exact Gerchberg-Saxton alternating projections with local convergence. 
If the row batchsize is relatively large this error may be small and  $\textbf{x}^+$ may well-approximate $\tilde{\textbf{x}}^+$, satisfying the local linear convergence conditions of Refs. \cite{Lewis2009LocalProjections, Kruger2016RegularityProjections}.
Conversely, if the row batchsize is very small, 
the deviation between $\textbf{x}^+$ and $\tilde{\textbf{x}}^+$ will grow large, 
which has  potential to disturb local convergence as projections become highly inexact.  

\begin{figure}[!t]
\centerline{\includegraphics[width=\columnwidth]{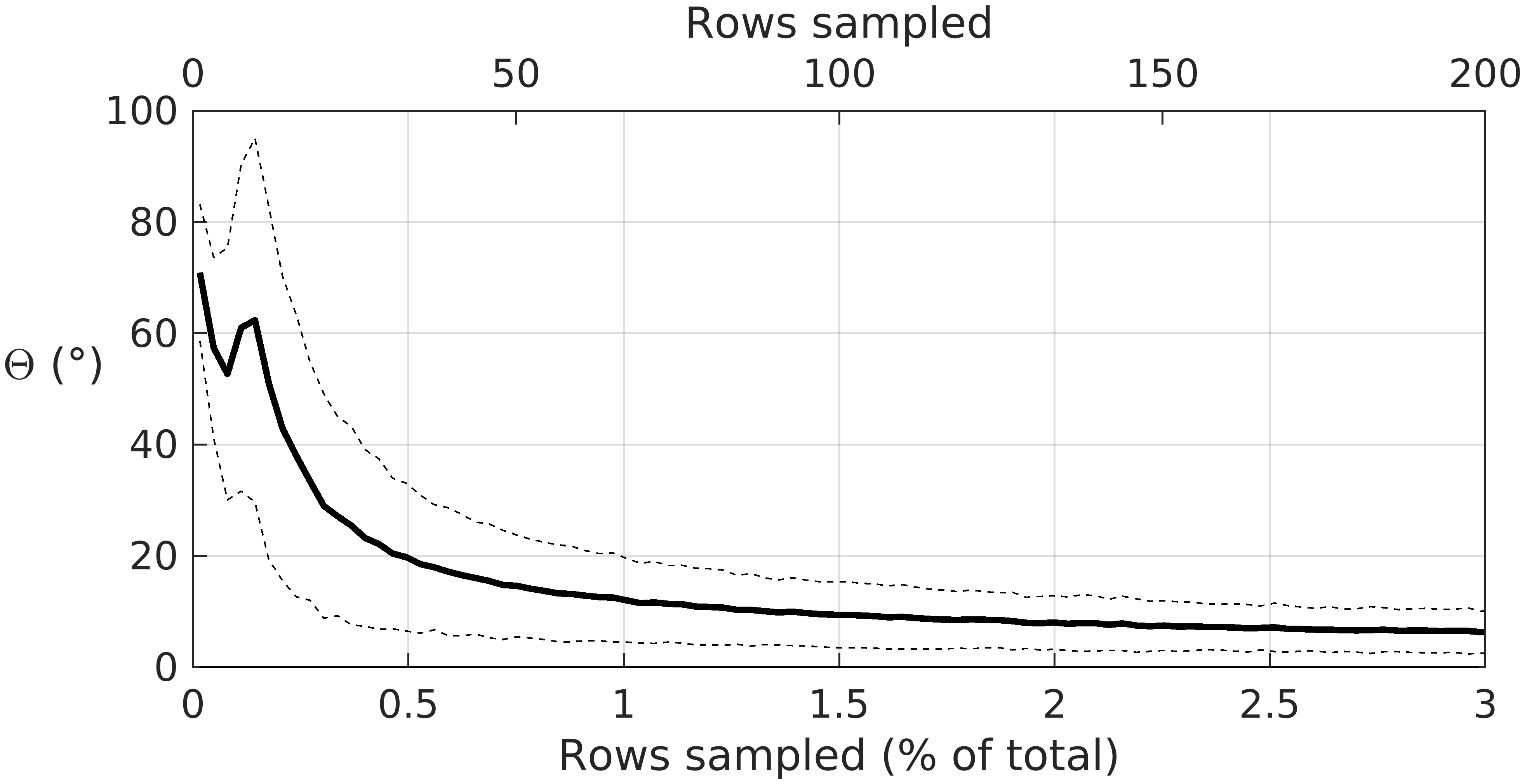}}
\caption{ { Mean angle between $\textbf{x}^+-\textbf{x}$ and $\tilde{\textbf{x}}^+-\textbf{x}$ over degree of row sampling of $\textbf{A}$, with 95\% confidence interval. At large batchsizes the angle $\theta$ is small; however, below a batchsize of approximately 0.5\% of the total row dimension orthogonality increases rapidly.}}
\label{fig1}
\end{figure}

\subsection{Minibatched GS Pulse Design Algorithms}
 A two-stage inexact Gerchberg-Saxton RF pulse design algorithm was implemented to solve the MLS pulse design problem, outlined in Algorithm \ref{alg:inexact}. 
 The goal of the algorithm was to balance global exploration of the various local minima of the MLS design problem with local convergence. 
 After initialization (Alg. \ref{alg:inexact}, lines 1-5), 
 the algorithm begins with an initial stage in which inexact alternating projections are interleaved with exact alternating projections  {(Alg. \ref{alg:inexact}, lines 6-15)}. 
 In this stage, inexact alternating projection iterations allow for global exploration by breaking the requirements for local convergence of the method of alternating projections, while exact alternating projection iterations provide local convergence and keep the projection sequence in the neighborhood of the set of feasible points of the phase retrieval problem. We note that the inclusion of iterations that violate the convergence guarantees of alternating projections implies that the final $(\textbf{x}_k, \textbf{z}_k)$ pair may not provide the lowest cost of the candidate points explored by the algorithm. As a result, this first stage of the algorithm returns the lowest-cost solution.

 \par The initial optimization stage is followed by a solution refinement stage (Alg. \ref{alg:inexact}, lines  {16-17}), in which the lowest-cost $(\textbf{x}_k, \textbf{z}_k)$  { with respect to Eqn \ref{mls}} is used to initialize a conventional Gerchberg-Saxton alternating projection sequence. 
 The inclusion of this final stage guarantees that the algorithm returns the lowest-cost solution ($\textbf{x}^*, \textbf{z}^*$) contained in the local convergence neighborhood $\mathcal{B}(\textbf{z}^*, \delta)$ in which the first stage finishes.
 Implementation and demonstration of Algorithm 2 can be found at our repository \url{https://github.com/jonbmartin/minibatch_MLS_demo}.

 \begin{algorithm}
    \captionof{algorithm}[Interleaved Inexact GS Pulse Design Algorithm (PseudoCode)]{Interleaved Inexact GS Pulse Design Algorithm}
    \begin{algorithmic}[1]
        \State \textbf{Input:} $\textbf{A} \in \mathbb{C}^{N_s \times N_cN_t}$, $\textbf{x}_0\in \mathbb{C}^{N_cN_t}$, $\textbf{b} \in \mathbb{C}^{N_s}$
        \State $k \gets 1, t\gets \text{inexact projection interval}$
        \State $r \gets \text{batchsize}$
        \State $\textbf{x}_k \gets \textbf{x}_0$
        \State $\textbf{z}_k \gets \angle (\textbf{Ax}_k)$

        \State Generate $N_{iter}/t$ row-sampled $\mathbf{\tilde{A}}$, $\mathbf{\tilde{y}}$
        
        \Comment{Begin Interleaved Inexact Stage}
        \For{k = 1 to $N_{iter}$}
            \State $\textbf{y} \gets \textbf{b} \cdot \textbf{z}_k $
            \If{\textit{k} mod t == 0}
            \State $\textbf{x}_k \gets (\mathbf{\tilde{A}}^H_k\mathbf{\tilde{A}}_k)^{\dagger}\mathbf{\tilde{A}}^H_k\mathbf{\tilde{y}}_k$
            \Else
                \State $\textbf{x}_k \gets (\textbf{A}^H\textbf{A})^{\dagger}\textbf{A}^H\textbf{y}$
            \EndIf

            \State $\textbf{z}_k \gets  (\textbf{Ax}_k)/|\textbf{Ax}_k|$

            \State Calculate cost of ($\textbf{x}_k, \textbf{z}_k$)

        \EndFor
        \State $\textbf{x} \gets \textbf{x}_k$ providing lowest cost  { w.r.t. Equation \ref{mls}}
        
        \Comment{Begin Exact Refinement Stage}
        \State $(\textbf{x}^*, \textbf{z}^*) \gets \text{Algorithm 1}(\textbf{A}, \textbf{x}, \textbf{b})$

        
        \State return $(\textbf{x}^*, \textbf{z}^*)$
        \end{algorithmic}
    \label{alg:inexact}
\end{algorithm}

 \par Algorithm \ref{alg:inexact} is appropriate for RF shimming or the design of 1-spoke RF pulses, or designing spectral-spatial excitation. 
 We also use this algorithm as a building block for a multi-spoke pulse designer, outlined in Algorithm \ref{alg:multispoke_inexact}. 
 Algorithm \ref{alg:multispoke_inexact} follows the interleaved greedy and local  multi-spoke pulse design Algorithm of Ref \cite{Grissom2012Small-tip-angleMethods.}, 
 but the inexact GS design algorithm is used to update the RF subpulse weights instead of conventional GS.  {This demonstrates how the proposed method can be a part of more complex, multi-stage pulse designs.}

\par
\begin{figure}[!t]

\centerline{\includegraphics[width=\columnwidth]{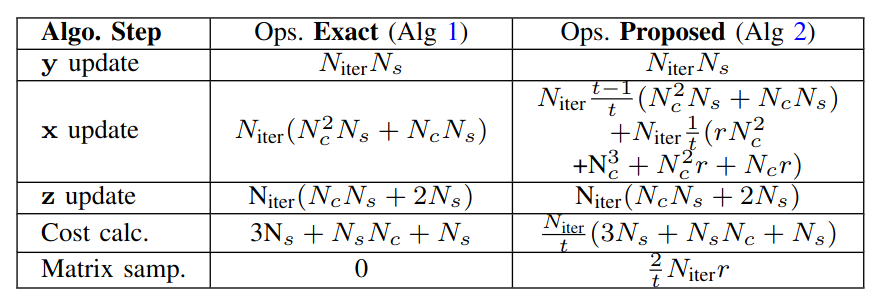}}
\caption{Table 1: Algorithm Operations Comparison}
\label{tab1}
\end{figure}

 Table \ref{tab1} shows a comparison of operations required for the exact GS algorithm (Alg. \ref{alg:GSalg}) and the interleaved inexact stage of Alg. \ref{alg:inexact}. 
Since they are identical, the $\textbf{y}$ and $\textbf{z}$ updates have the same number of operations for the two algorithms. 
The $\textbf{x}$ update in Alg. \ref{alg:inexact} uses the modified inexact update every $t$ iterations. 
Overall, this results in a computational savings for most problem sizes, 
since the increased number of cost calculations and matrix sampling is much smaller than the savings produced by the smaller dimensions of the matrix multiplications in Alg. \ref{alg:inexact}. 
For example, consider an optimization with $N_s$ = 5000, $N_c$ = 8, $t = 3$, $N_{iter}$ = 100, $r = 10$. 
To update $\textbf{x}$ in the exact Alg. \ref{alg:GSalg} one can expect approximately $3.6\times10^7$ operations, 
while for the proposed Alg. \ref{alg:inexact} one expects $2.4\times10^7$ operations, a 1/3 reduction. 
This reduction is far greater than the approximately $2\times10^6$ operations required to regularly calculate cost or the small number required for matrix sampling. 
Later in this work, we will further show that in general, 
for the same number of iterations the new algorithm also yields a lower-cost solution.

\begin{algorithm}
    \captionof{algorithm}[Inexact Multispoke Design Algorithm (PseudoCode)]{Inexact Multispoke Design Algorithm}
    \begin{algorithmic}[1]
        \State $N_{spk} \gets 1$
        \State $\mathbb{K} \gets \{(0,0)\} \{\text{Start with a DC spoke.}\}$
        \State Construct $\textbf{A} \in \mathbb{C}^{N_s \times N_cN_{spk}}$
        \State $\textbf{b} \gets$ target magnetization
        \While{$N_{spk} <$  $N_{spk, max}$}
            \While{Cost criterion not met}
                \State $(\textbf{x}^*, \textbf{z}^*) \gets \text{Algorithm 2(\textbf{A},  \textbf{b})}$ ($\mathbb{K}$ fixed)
                \State Update $\mathbb{K}$ ($\textbf{x}^*, \textbf{z}^*$ fixed)
                \State Evaluate cost
            \EndWhile
            \State $N_{spk} \leftarrow N_{spk} + 1$
            \State Determine new spoke $(\tilde{k}^x, \tilde{k}^x)$, append to $\mathbb{K}$
            \State Build new $\textbf{A} \in \mathbb{C}^{N_s \times N_cN_{spk}}$
        \EndWhile
    \State return $(\textbf{x}^*, \textbf{z}^*), \mathbb{K}$        \end{algorithmic}
    \label{alg:multispoke_inexact}
\end{algorithm}

\section{Methods}
The following sections describe an empirical investigation into the conditions in which using an inexact alternating projection provides exploration of the Gerchberg-Saxton shimming problem. Numerical investigations into the sampling of the system matrix in the context of RF pulse design are presented first.
Characterization of the algorithm was then performed in simulation and in vivo validation was performed on a 7T human MRI scanner. 

\subsection{Numerical Investigation of Inexact Gerchberg Saxton Alternating Projection in RF Shimming}
Two numerical examples were examined to characterize the effect that subsampling the row dimension of the MLS system has on the projection of Eqn \ref{projection1} in an RF pulse design context. 
The objectives of these investigations were to 1) determine the  { distribution of angles between vectors $\textbf{x}^+-\textbf{x}$ and $\tilde{\textbf{x}}^+-\textbf{x}$ produced by performing updates with the minibatched system matrix}, 
and 2) show the changes in the  { linear least squares (LLS) problem} cost surface to demonstrate that inexact projections can temporarily increase cost, breaking local convergence.
For the examples considered, $\textbf{A} \in \mathbb{C}^{N_s \times N_c}$ was formed from the central axial slice of simulated and in-vivo 8-channel and simulated 30 channel $B_1^+$ maps in the human head at 7T. 
These simulated $B_1^+$ maps were produced by the procedure outlined in the Methods section III.B. The simulated $B_1^+$ maps had 6,247 sampled spatial locations, creating a row dimension in $\textbf{A}$ of 6,247 from which sample sets were drawn $(\mathbf{\tilde{A}} \in \mathbb{C}^{6,247 \times N_c})$. 
The 8-channel in-vivo maps had 2,238 sampled spatial locations,  
giving $(\mathbf{\tilde{A}} \in \mathbb{C}^{2,238 \times N_c})$. 
They were produced using the experimental method outlined in Section III.C. 

\subsubsection{Variation in Projection Direction with Batchsize}
Using the 8-channel simulated $B_1^+$ maps, 
1,000 random $(\textbf{x}, \textbf{z})$ complex initialization pairs were generated by drawing the real and imaginary components of the $i$th $\textbf{x}$ from the standard normal distribution, 
and computing $\textbf{z}_i = (\textbf{Ax}_i)/|\textbf{Ax}_i|$. 
In each trial, one step of the LLS projection (Equation \ref{projection1}) was taken with either an exact $\textbf{A}$, 
giving $\textbf{x}^+ = \textbf{A}^\dagger (\textbf{b} \cdot \textbf{z})$, 
or a row-sampled system $\tilde{\textbf{A}} = \textbf{SA}$, 
giving $\tilde{\textbf{x}}^+= \left(\textbf{SA}\right)^\dagger \left(\textbf{S}\left(\textbf{b} \cdot \textbf{z}\right)\right)$. 
This was repeated across batchsizes in the range $[1:1:175]$. For each trial $i$, the vector angle $\theta$ between the exact update $\textbf{x}^+-\textbf{x}$ and the inexact update $\tilde{\textbf{x}}^+-\textbf{x}$ was calculated to measure the deviation between the exact update and the inexact update. 




\subsubsection{Cost Surface Changes with Batchsize}
\begin{figure*}[!t]
\centerline{\includegraphics[width=\textwidth]{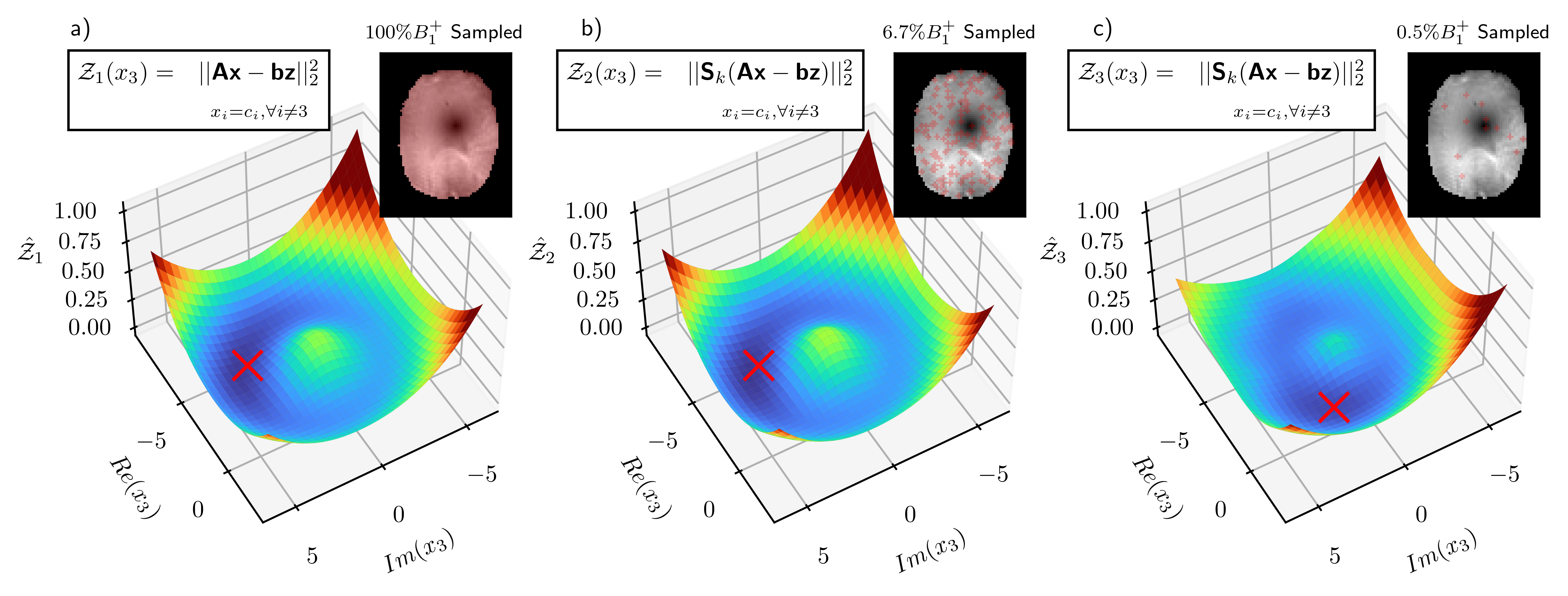}}
\caption{Cost surfaces for the LLS optimization of Equation \ref{minibatched_optimization} across a single channel's complex weights. The red crosses on the plots mark locations of minimum cost. 
a) shows the case of full batch sampling (exact GS), b) shows sampling $6.7\%$ of the row dimension, and c) shows sampling $0.5\%$ of the row dimension. 
The inset plots of the slice's magnetization slice show the current value of $|\textbf{Ax}|$, 
with spatially sampled locations in each case marked in red. 
There is little difference between the cost surfaces in a) and b), 
indicating that many measurements can be removed from $\textbf{A}$ without shifting the problem's minima. However (c) shows that the cost surface was reshaped when only a few rows were sampled.}
\label{fig3}
\end{figure*}

Shimming of the in-vivo 8 channel 7T slice was run to convergence using the conventional exact GS algorithm (Alg \ref{alg:GSalg}) 
with a circularly polarized initializer, 
where the circularly polarized initializer was achieved by aligning the $B_1^+$ fields' phase in the centroid of each slice, 
with equal amplitude weights on each channel.
This produced a solution $(\textbf{x}^*, \textbf{z}^*)$, 
which resulted in a signal null in the excitation pattern $\textbf{m} = |\textbf{Ax}^*|$ (see Fig. \ref{fig3} inset). 
Three $\textbf{A}$ system matrices were built from the $B_1^+$ maps: one with all spatial samples included $(\textbf{A} \in \mathbb{C}^{2,238\times8})$, one with 6.7\% of rows sampled $(\textbf{A} \in \mathbb{C}^{150\times8})$, and one with 0.5\% of rows sampled $(\textbf{A} \in \mathbb{C}^{12\times8})$. 
For each system matrix, the LLS cost of $\textbf{x}$ candidate shims was evaluated in the range [-7, 7] for each real and imaginary shim weight. 

\subsection{Simulations}

To evaluate the GS algorithm, 
two 10-subject datasets (5M, 5F) of 7T $B_1^+$ maps were generated via simulation of an 8-channel and 30-channel loop coil in HFSS (Ansys Inc., Canonsburg, PA, USA). 
Both arrays had 28 cm diameters to accommodate a receive coil insert, 
with heights of 16 cm and 17.8 cm for the 8-channel and 30-channel coils, respectively. 
Simulations were produced with 20 $\times$ 20  $\times$ 14 cm\textsuperscript{3} FOV and 2 mm isotropic voxel size \cite{Grissom2019}. 
To simulate the effect of designing shims using $B_1^+$ maps of different realistic resolutions, 
the 8-channel simulated $B_1^+$ maps were interpolated to a finer in-plane grid of 1.5 $\times$ 1.5 $\times$ 2.0 mm$^3$ voxels and 1.0 $\times$ 1.0 $\times$ 2.0 mm$^3$ voxels. 

\subsubsection{Simulation I: Optimal Minibatch Size}
The first simulation evaluated the optimal batchsize for shimming a full head slice-by-slice using Algorithm \ref{alg:inexact}. 
Algorithm \ref{alg:inexact} was run to design 2D slice-specific RF shims for axial multislice 2D imaging over batchsizes of $[1:1:100]$  for the 8 channel dataset across the three $B_1^+$ acquisition resolutions and batchsizes of  $[1:1:200]$ for the 30 channel datasets. 
A total of 60 unique $B_1$ map slices were each shimmed 100 times in each dataset. 
A set of 100 random complex initializers pulled from the standard normal distribution were used to initialize $(\textbf{x}, \textbf{z})$. 
The algorithm was run for 350  iterations of the initial inexact stage, 
with an inexact projection interval of $t = 3$. The final refinement stage of  Alg \ref{alg:inexact} was run for 15 iterations to reach local convergence. 
Shimming was repeated with the exact GS Algorithm \ref{alg:GSalg} using the same initializers; 
the exact GS algorithm was stopped after 350 iterations or when the algorithm reached a difference in cost between successive iterations of $\epsilon < 10^{-5}$. 

\begin{figure*}[!t]
\centerline{\includegraphics[width=\textwidth]{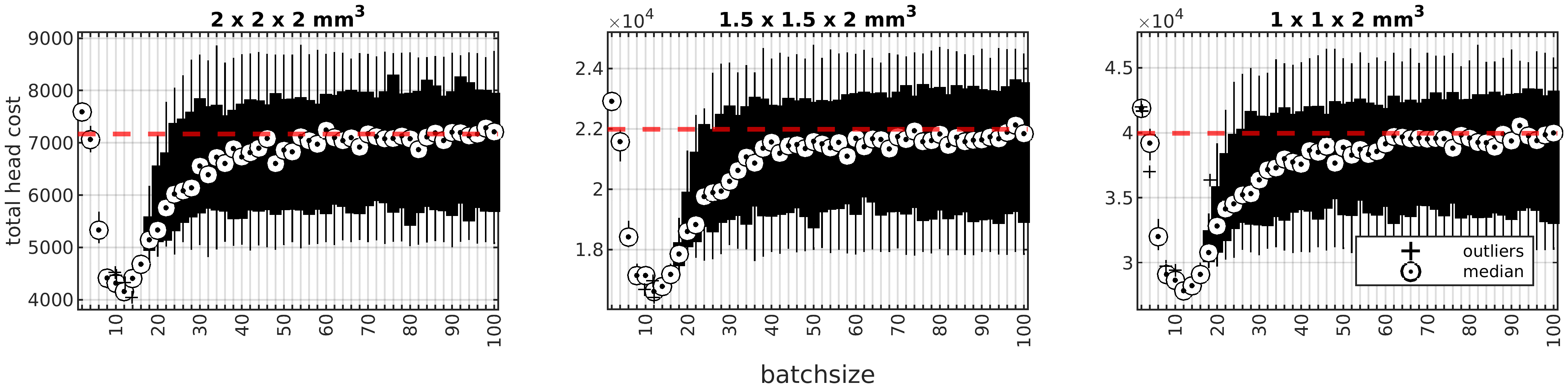}}
\caption{ { Simulation I: Optimal Minibatch Size (8-element coil).} Mean total head cost produced by the inexact GS algorithm as a function of batchsize, for 100 random initializations, across three different in-plane resolutions of 8-channel $B_1^+$ maps. The red horizontal line shows the mean cost of conventional exact GS for the same set of random initializations. At small batchsizes there is a large reduction in mean and variance of total head cost across all three $B_1^+$ datasets. As batchsize is increased the variance increases and the mean cost gradually converges to the exact GS mean. At very small batchsizes ($\approx \leq8$), mean cost spikes sharply.} 
\label{fig4}
\end{figure*}

\subsubsection{Simulation II: Optimal Interleaved Exact GS Iterations}

A study was conducted to determine the optimal number of exact GS iterations to interleave between inexact GS iterations for small batchsizes. 
For a fixed batchsize of $r = 12$, a full axial 8-channel head dataset was shimmed with 100 random initializations following Algorithm \ref{alg:inexact}. 
This was repeated with inexact iteration interval $t = [0:1:4]$. 
In each case, shimming was run for each slice until 2500 exact GS iterations had been performed. 
Randomly initialized exact GS, the conventional method, was also performed with the same set of random initializations. 


\subsubsection{Simulation III: 1-Spoke and 2-Spoke L-Curves}
The feasibility of using the interleaved inexact GS algorithm, 
Algorithm \ref{alg:multispoke_inexact}, 
to design power-regularized 1-spoke and 2-spoke RF pulses was investigated. 
Using a central axial slice of one head of the simulated 8-channel axial $B_1^+$ dataset, 
1-spoke and 2-spoke pulses were designed with 50 power regularization values evenly distributed on a logarithmic scale in the range $\beta/N_s = \left[1 \times 10^{-9}, 1 \times 10^{1}\right]$ for the 2-spoke design and $\beta/N_s = \left[1 \times 10^{-8}, 1 \times 10^{-5}\right]$ for the 1-spoke design, for $B_1^+$ maps with a maximum of 4.17 uT. 
For each $\beta$ value, 100 trials were conducted with random initialization. 
The algorithm was run for 350 iterations, 
with $t=3$ interleaved inexact iteration interval and with 10 exact GS iterations initialized by the minimum-cost solution following the interleaved inexact optimization. 

\subsubsection{Simulation IV: Population Performance}
To verify the robustness of the algorithm across different human subjects, 
the algorithm was tested across the full 10-head 8-channel dataset, 
for slice-by-slice shimming in each of the axial, sagittal, and coronal dimensions. 
For each slice in each head, optimization was performed with conventional exact GS optimization, 
interleaved inexact GS optimization (Algorithm \ref{alg:inexact} with $t=3$), 
and fully inexact GS optimization (Algorithm \ref{alg:inexact} with $t=1$, 
using an inexact projection in all iterations). 
All inexact iterations used a batchsize of 12 for all slices. 
100 trials of 350 full algorithm iterations were performed with random initialization, 
with 10 exact GS iterations initialized by the minimum-cost solution following each optimization. 

\subsubsection{Simulation V: Spectral-Spatial 3T RF Pulse Design}
 {  To evaluate the proposed algorithm 
in a de novo MLS RF pulse design problem,
it was applied in simulation to the design of spectral-spatial selective RF pulses for water-selective imaging. 
The simulation used a dataset of in-vivo human abdominal 3T 8-channel $B_1^+$ and $B_0$ maps used in Ref. \cite{MalikWater2010}.
Following the experiments of Ref. \cite{MalikWater2010}, 
sets of pulses were designed to provide a $20^\circ$ excitation at the 3T water frequency and $0^\circ$ excitation at the 3T fat frequency ($\Delta f \approx 435 $Hz) across a single abdominal imaging slice. 
These two frequencies of interest were weighted 1 and all others were weighted 0. 
The designed pulses included 1) the unoptimized 1-3-3-1 binominal pulse used with the transmit coil driven in quadrature mode used in \cite{MalikWater2010}, 2) an 8-channel, 4-subpulse, time-bandwidth product 4 RF pulse optimized with the standard MLS GS method used in the original paper \cite{MalikWater2010}, 
and 3) an 8-channel, 4-subpulse pulse optimized with the proposed inexact minibatched MLS GS method in Algorithm \ref{alg:inexact}. 
For this experiment, an inexact GS batchsize of 1.5 times the column dimension of the pTx design matrix was used (batchsize = 48), 
as this was shown to provide approximately optimal performance in all other experiments. 
An inexact iteration interval of $k$=3 was used, and a refinement stage with 15 exact GS iterations was performed using the best result from the inexact stage. 
Both iterative methods were initialized with 100 random initializers, 
and were allowed to progress for 350 iterations in each trial.
These simulations were performed in MATLAB (MathWorks, Natick MA, USA) using the original authors' online code repository.}

\subsection{Scanner Experiment}
The interleaved inexact GS algorithm was used to design RF shim weights that were deployed on a  {Philips 7T Achieva system (Philips Healthcare, Best, Netherlands)}. 
The experimental methods followed the procedure used in Ref. \cite{Paez2021RobustRegularization}. 
Human-brain $B_1^+$ maps were acquired in a healthy 28 year old male volunteer using the 2D-multislice DREAM method \cite{Nehrke2012DREAMaMapping} with a $224 \times 224 \times 123 $mm$^3$ FOV, with 3.5 mm$^3$ isotropic resolution over 32 axial slices with a 0.35 mm slice gap. 
Three sets of RF shims were designed from these $B_1^+$ maps. 
Fixed CP-mode shims were set for each slice, by aligning the $B_1^+$ fields' phase in the centroid of each slice, with equal channel amplitude weights.
Exact GS shimming was performed for all 32 slices, 
using a CP initialization and iterating until 350 iterations were reached or $\epsilon \leq 1\times10^{-5}$. 
Inexact GS shimming was performed for the same slices, 
using a CP initialization and iterating through 350 iterations of the inexact GS algorithm with $t = 3$, batchsize = 12, and refining the lowest-cost $(\textbf{x},\textbf{z})$ with 15 exact GS iterations. 

\par  { To evaluate the design time, designs were repeated 100 times for each slice offline, after scanning was completed. Designs were performed with minibatched inexact GS with batchsize = 12 and a broader range of inexact iteration intervals were considered, $t=[1:1:4]$. This was compared to the computation time of exact GS. All designs were performed on a laptop CPU with Intel i9-13900HX processor (Intel Corporation, Santa Clara, CA, USA).}

\begin{figure}[!t]
\centerline{\includegraphics[width=\columnwidth]{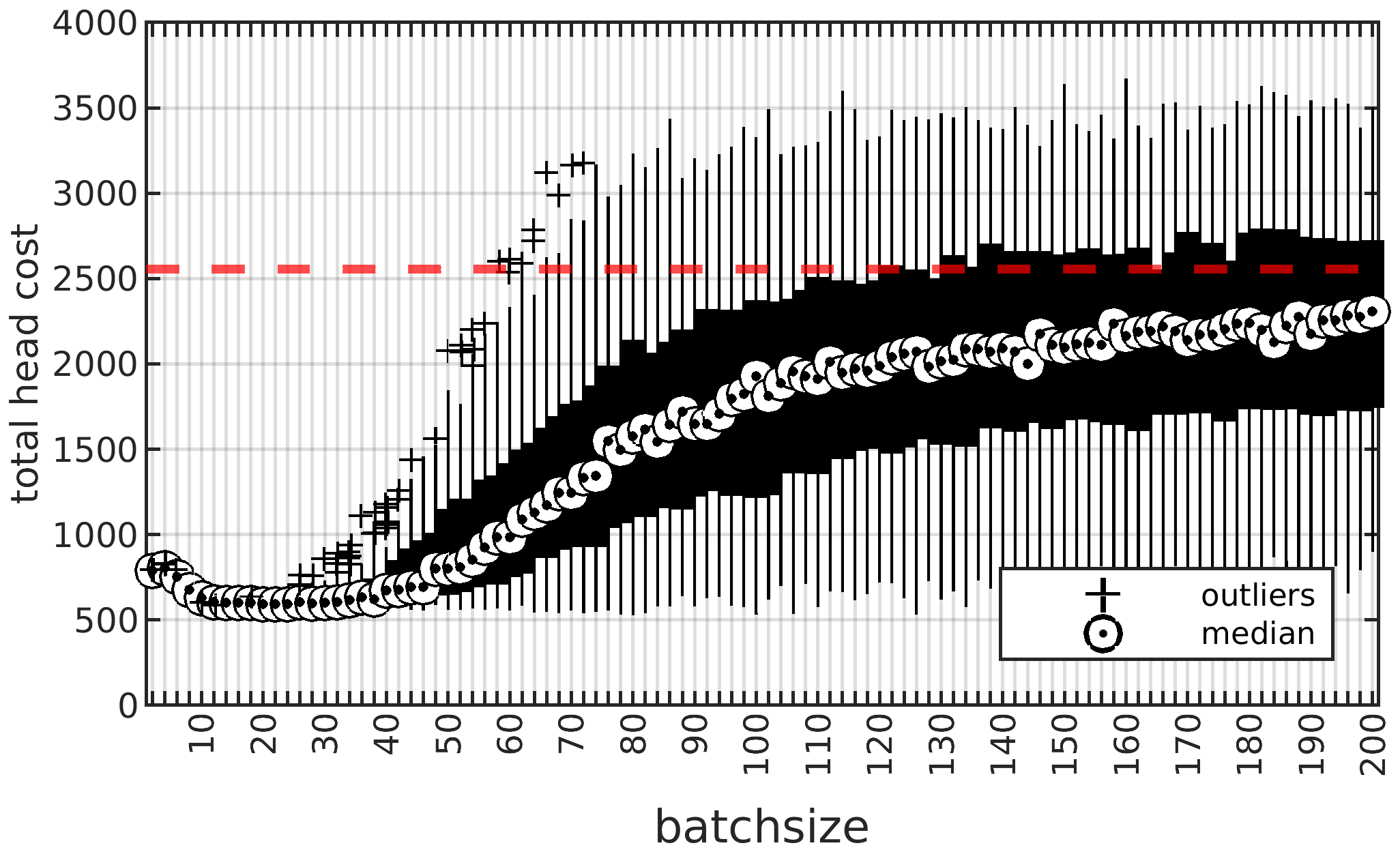}}
\caption{ { Simulation I: Optimal Minibatch Size (30-element coil).} Mean total head cost produced by the inexact GS algorithm as a function of batchsize, for 100 random initializations, for 30-channel $B_1^+$ maps. The red horizontal line shows the mean cost of conventional exact GS for the same set of random initializers. As in Figure \ref{fig4}, there is a decrease in mean head cost and variance with decreasing batchsize. A small increase in cost occurs in both cases for very small batchsizes, less than approximately 8 rows.}
\label{fig5}
\end{figure}

\par The RF shims were deployed in a single-shot GRE-EPI scan with dynamic slice-by-slice RF shimming, 
with $65^\circ$ FA, TR/TE = 2000/28 ms, 2.2 mm$^3$ isotropic resolution over a $220 \times 220 \times 79$ mm$^3$ FOV. 
The 32 designed shims were interpolated to shim weights for 36 interleaved slices with a 0-mm slice gap following the procedure of \cite{Paez2021RobustRegularization} in order to reduce shim design time.

\section{Results}
\subsection{Numerical Investigations}
\subsubsection{Variation in Projection Direction with Batchsize}
Fig. \ref{fig1} shows the average angle between $\textbf{x}^+-\textbf{x}$ and $\tilde{\textbf{x}}^+-\textbf{x}$ over different batchsizes of the rows of $\textbf{A}$. At large batchsizes, the angle $\theta$ is relatively small. However, as fewer rows of $\textbf{A}$ are used in the computation of $\tilde{\textbf{x}}^+$, the deviation between the exact update and the inexact update grows large, quickly approaching $90^\circ$ at batchsizes of $<0.5\%$. We can anticipate that the growing orthogonality between the average minibatch update and the batch update with decreasing batchsize will mean that the second condition of Eqn \ref{inexact_projection} will require larger values of $\sigma$ to define the inexact projection.  { Ref. \cite{Kruger2016RegularityProjections} }shows that there are upper bounds on the size of $\sigma$ for error reduction and convergence to be guaranteed. 


\subsubsection{Cost Surface Changes with Batchsize}
Figure \ref{fig3}a shows the LLS cost surface of this solution for the full system $\textbf{A}$ as a function of the complex weight of a single transmit channel, 
$\pazocal{Z}(x_3)$. 
Figure \ref{fig3}b shows the same cost surface but for a row-sampled 
version of the LLS problem, in which the system matrix $\textbf{A}$ is replaced by $\textbf{S}_k\textbf{A}$. 
The function of $\textbf{S}_k$ in this example is to subsample the rows of $\textbf{A}$. 
In this example 150 ($6.7\%$) of the system matrix's 2,216 rows were retained. 
For this level of subsampling, 
from Figure \ref{fig1} it is expected that there will not be a large deviation the subsampled update $\tilde{\textbf{x}}^+ - \textbf{x}$ versus the fully sampled update $\textbf{x}^+ - \textbf{x}$ (an average deviation angle $\theta$ of $\leq 10\%$). 
Indeed there is very little change in the cost surface despite a large reduction in the problem size. 
However, Figure \ref{fig3}c shows that subsampling the LLS system to a very small number of rows
significantly changes the location of the local minima.

\begin{figure}[!t]
\centerline{\includegraphics[width=\columnwidth]{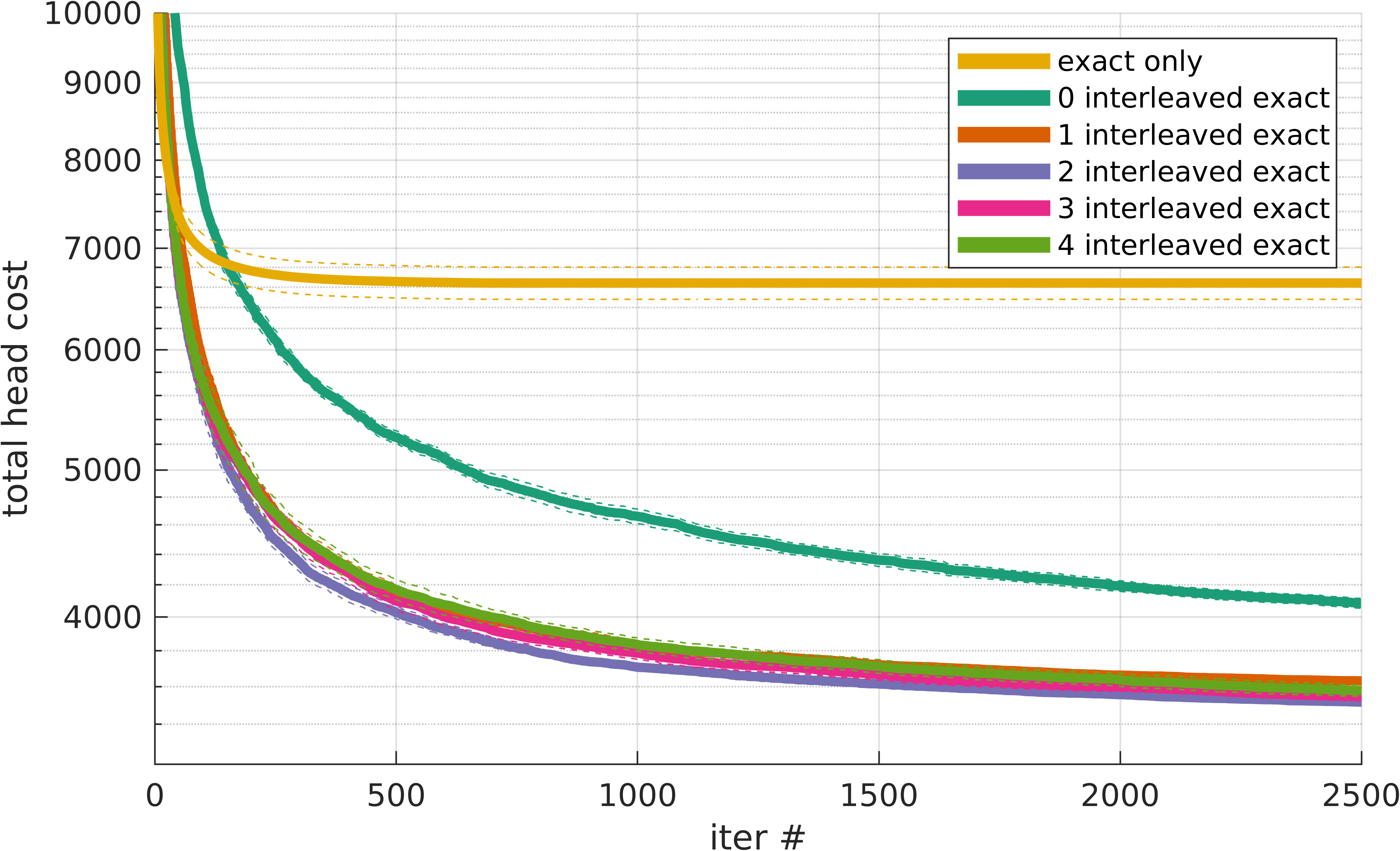}}
\caption{ { Simulation II: Optimal Interleaved Exact GS Iterations.} Cost reduction versus number of exact GS iterations interleaved between inexact GS iterations. Inexact GS with 0 interleaved exact GS iterations is shown for comparison (`0 interleaved inexact').
Solid lines show mean lowest cost at each iteration, and dotted lines show 95$\%$ CI. Using inexact projections produces improvements in cost and avoids premature convergence. Interleaving any number of exact GS iterations considered dramatically improves the efficiency of the algorithm. Interleaving 2 exact GS iterations (purple) results in the most efficient cost reduction in this example. }
\label{fig4d8}
\end{figure}


\subsection{Simulations}
\subsubsection{Simulation I: Optimal Minibatch Size}

Figure \ref{fig4} shows the distribution of total head cost across batchsizes for the three resolutions of 8-channel $B_1^+$ maps. The distribution varies with batchsize in approximately the same way for all three $B_1^+$ resolutions. 
For large batchsizes (approximately 40 and higher), 
the mean of the total head cost is approximately equal to the mean total head cost for exact GS optimization. 
In the batchsize range of approximately 8 -- 30 there is a significant reduction ($p < 0.005$) in mean total head cost and a large 
reduction in the variance in total head cost. 
At very small batchsizes (less than 8) the mean total head cost rose sharply and became larger than that of exact GS, 
which is likely due to the increase in condition number for these 
sizes as the number of independent rows included decreased sharply.
The curves show that that costs do not 
depend strongly on the spatial resolution of the $B_1^+$
maps, reflecting the fact that 
the electrodynamics of the $B_1^+$ fields on which the multi-channel shim/pulse design problem are based are captured across all these scales.

\par Figure \ref{fig5} shows the distribution of total head cost across batchsizes for the 30-channel coil. 
Again, the total mean cost converges towards the exact GS cost at large batchsizes, 
although this occurs more slowly than in the 8-channel case. 
For the 30-channel system, there is only a small increase in cost at very small batchsizes, 
and a larger range of batchsizes over which a cost reduction is reached.


\subsubsection{Simulation II: Number of Interleaved Exact GS Iterations}
Figure \ref{fig4d8} shows the convergence of the algorithm across number of interleaved exact GS iterations. Number of exact iterations are considered since these are the primary computational cost of the inexact GS algorithm (see Table \ref{tab1}).
Interleaving any number of exact GS iterations in the range [1, 4] produces large improvements in cost reduction versus using exclusively inexact GS iterations. 
Interleaving 2 exact iterations between inexact iterations reduces cost most efficiently, in terms of number of exact iterations required to reach a given cost. 
Interleaving additional exact iterations gradually reduces the efficiency of the algorithm, as this prioritizes local convergence without sufficiently regular global exploration.


\begin{figure}[!t]
\centerline{\includegraphics[width=\columnwidth]{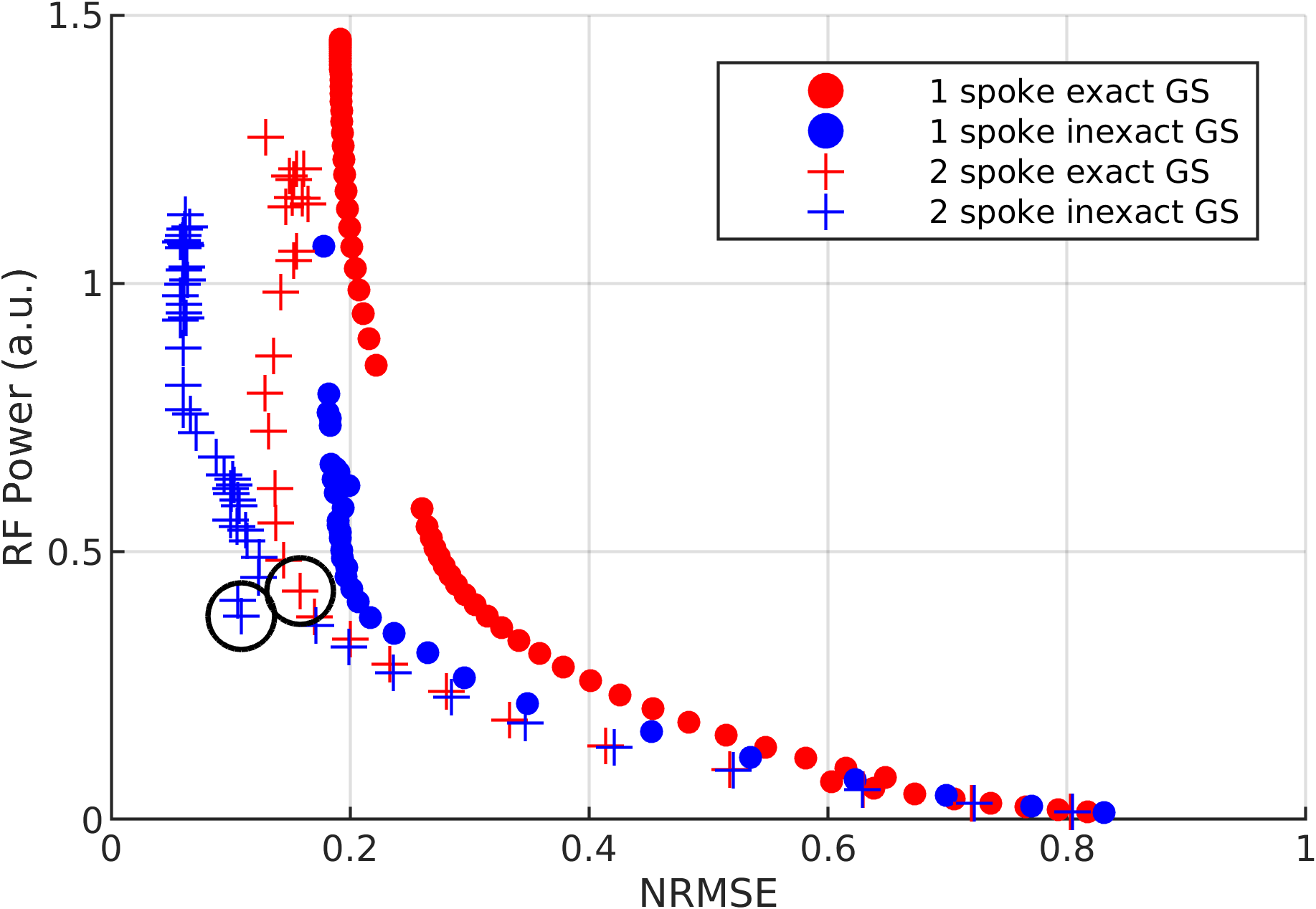}}
\caption{ { Simulation III: 1-Spoke and 2-Spoke L-Curves.} L-curve for 1-spoke and 2-spoke exact and inexact GS RF pulse design, with random initialization. Scatter points represent mean power and NRMSE values across trials. Inexact GS shifts the l-curve towards the origin in the case of both 1- and 2-spoke pulse design, robustly producing lower cost across a wide range of regularization values. Circled scatter points are the lowest-cost 2-spoke solutions ($\beta/N_s =5.72 \times 10^{-4}$). }
\label{fig4d10}
\end{figure}

\begin{figure}[!t]
\centerline{\includegraphics[width=\columnwidth]{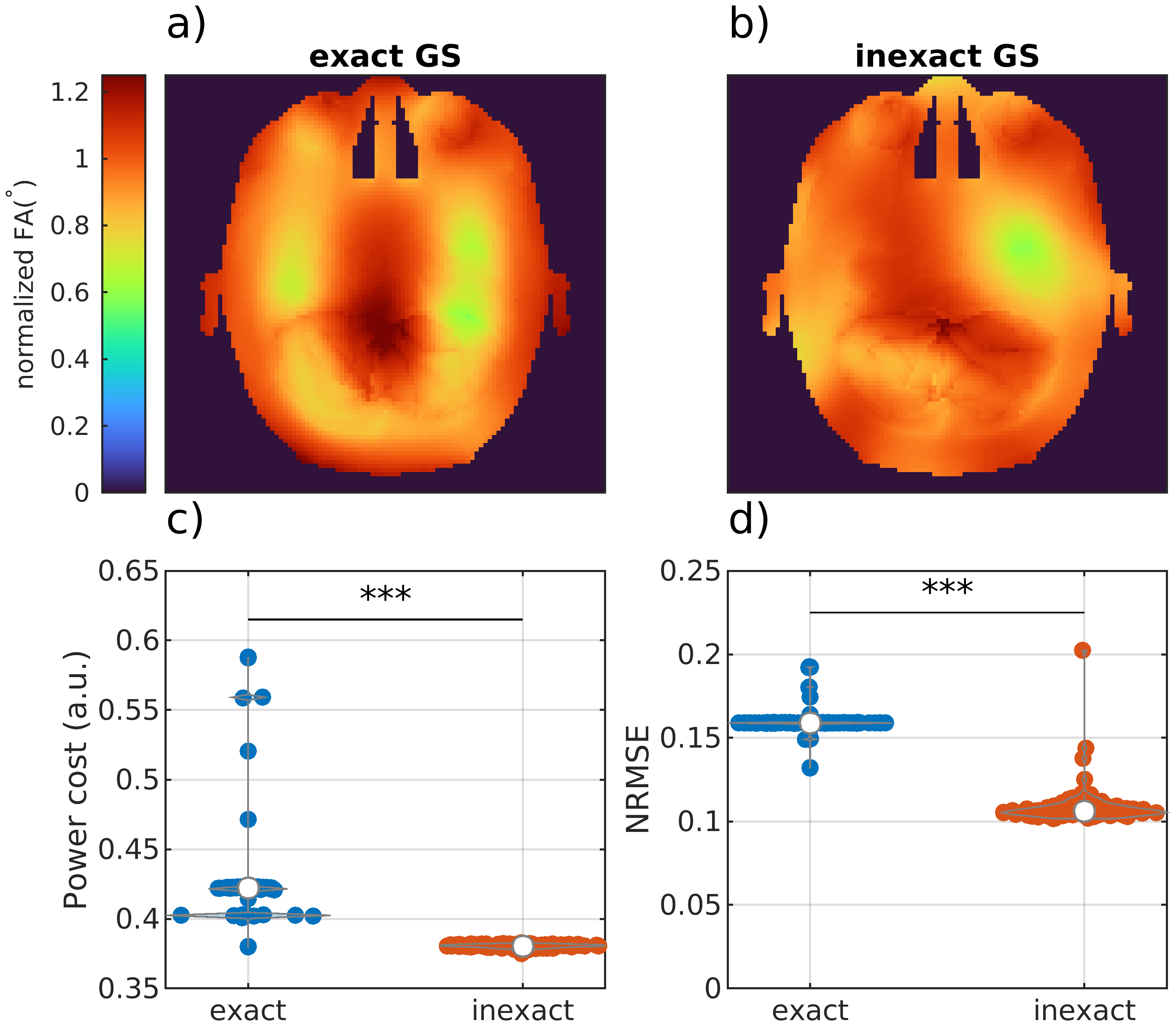}}
\caption{ { Simulation III: 1-Spoke and 2-Spoke L-Curves.} Median-error magnitude excitation  { in normalized flip angle} for the exact 2-spoke GS method a) and interleaved inexact method b) with random initialization with $\beta/N_s =5.72 \times 10^{-4}$ (circle, Fig \ref{fig4d10}).  { Figures (c) and (d) show violin plots of solution costs.} Significant (*** = $p < 0.001$) reductions in mean RF power cost (c) and error (d) were achieved using the inexact interleaved GS method. 
Exact GS produces several high-power outliers, while inexact GS produces none.}
\label{fig4d11}
\end{figure}


\subsubsection{Simulation III: 1-Spoke and 2-Spoke Design L-Curves}
Figure \ref{fig4d10} shows the l-curves produced by the varying power regularization parameter $\beta$. 
In both the 1-spoke (shimming) and 2-spoke cases, using inexact GS finds a more favorable power versus error tradeoff than exact GS for most regularization values, given the same budget of iterations. 
Figure \ref{fig4d11} shows the 2-spoke results for the regularization value that produced the lowest total cost ($\beta =5.72 \times 10^{-4}$) for the inexact 2-spoke method, 
which was the lowest total cost of the four spokes designs. 
Inhomogeneities are present in both magnitude shims due to the tradeoff of increased excitation error for reduced power. 
However, in both cases inexact 2-spoke GS produces a highly significant reduction in both power cost and NRMSE. 
For this regularization value, the exact GS 2-spoke solutions cluster at a few high cost minima across random initializations. Inexact GS solutions have fewer outliers and lower median cost. 

\begin{figure}[!t]
\centerline{\includegraphics[width=\columnwidth]{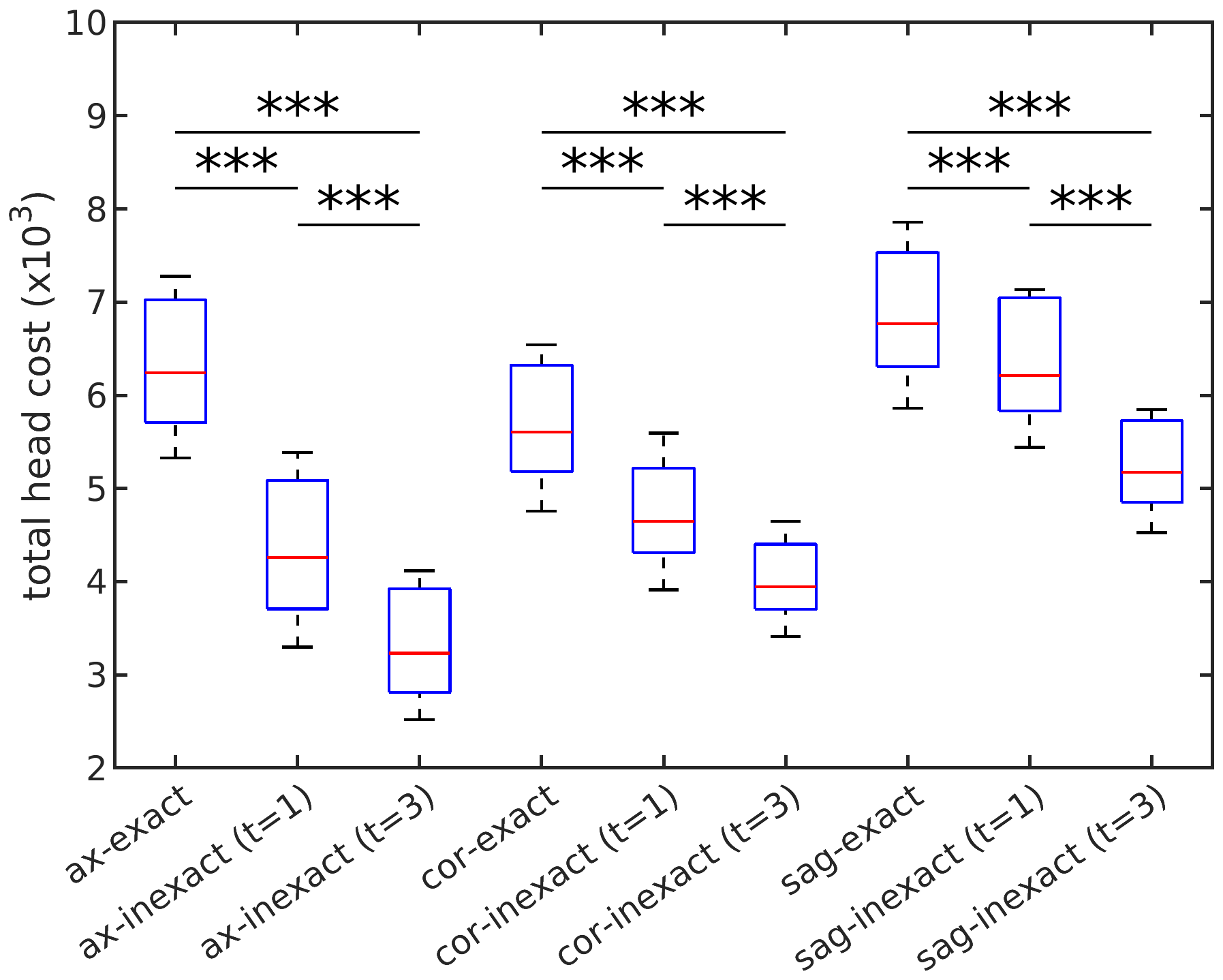}}
\caption{ { Simulation IV: Population Performance.}  { Box-and-whisker plot of} population results from shimming all 10 heads in the 8-channel axial 2x2x2 mm$^3$ dataset slice-by slice with the exact GS method, inexact GS with batchsize = 12 and $t=1$ (no interleaved exact GS iterations), and inexact GS with $t=3$ (2 interleaved exact GS iterations). Across all three axes, using either inexact GS approach results in a highly significant (*** = $p < 0.001$) reduction in the mean head cost.}
\label{fig4d13}
\end{figure}

\subsubsection{Simulation IV: Population Performance}
Figure \ref{fig4d13} shows the distribution of total full-head cost across the 10-head 3-axis 8-channel coil dataset, 
shimmed slice-by-slice in 100 randomly-initialized trials. 
Results are shown for conventional exact GS, inexact GS with no interleaved exact iterations ($t=1$, `inexact'), 
and inexact GS with 2 interleaved exact iterations, ($t=3$, `interleaved'). 
Across all three geometries (axial, sagittal, and coronal shimming), inexact GS with no interleaved exact iterations produces a highly significant reduction in mean total head cost. 
Interleaving exact GS iterations produces a highly significant reduction in mean total head cost versus both exact GS and inexact GS with no interleaved exact iterations, providing the lowest overall error in all three geometries. 

\begin{figure}[!t]
\centerline{\includegraphics[width=\columnwidth]{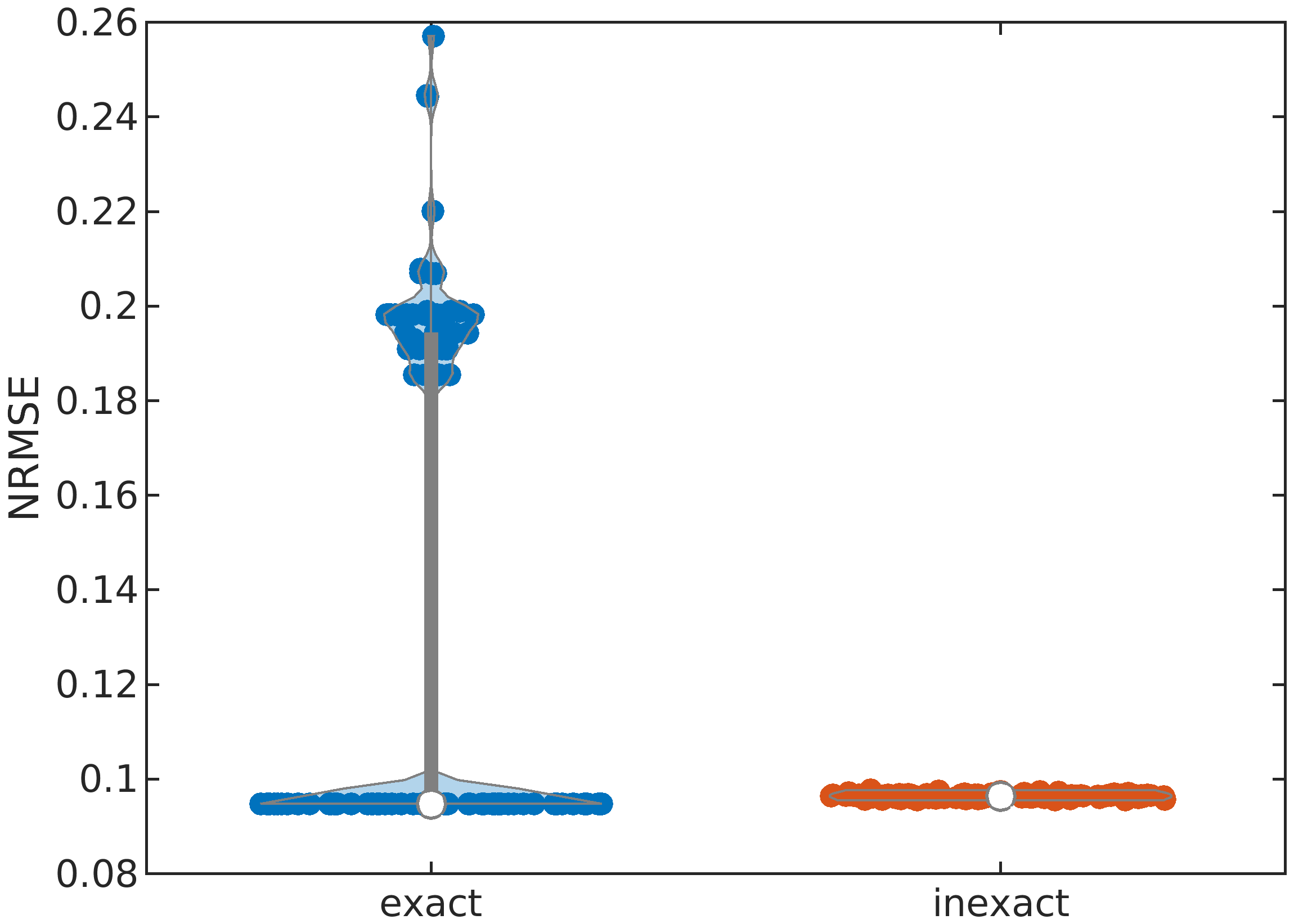}}
\caption{ { Simulation V: Spectral-Spatial 3T RF Pulse Design. Violin plots of design NRMSE across 100 trials for exact and inexact GS. Individual trial results are overlaid, and the median solution is plotted with a white circle. Median solution costs are similar, but exact GS has a large number of high-error failures (38\% of all trials). }}
\label{fig_spec_spat_violin}
\end{figure}

\begin{figure*}[!t]
\centerline{\includegraphics[width=\textwidth]{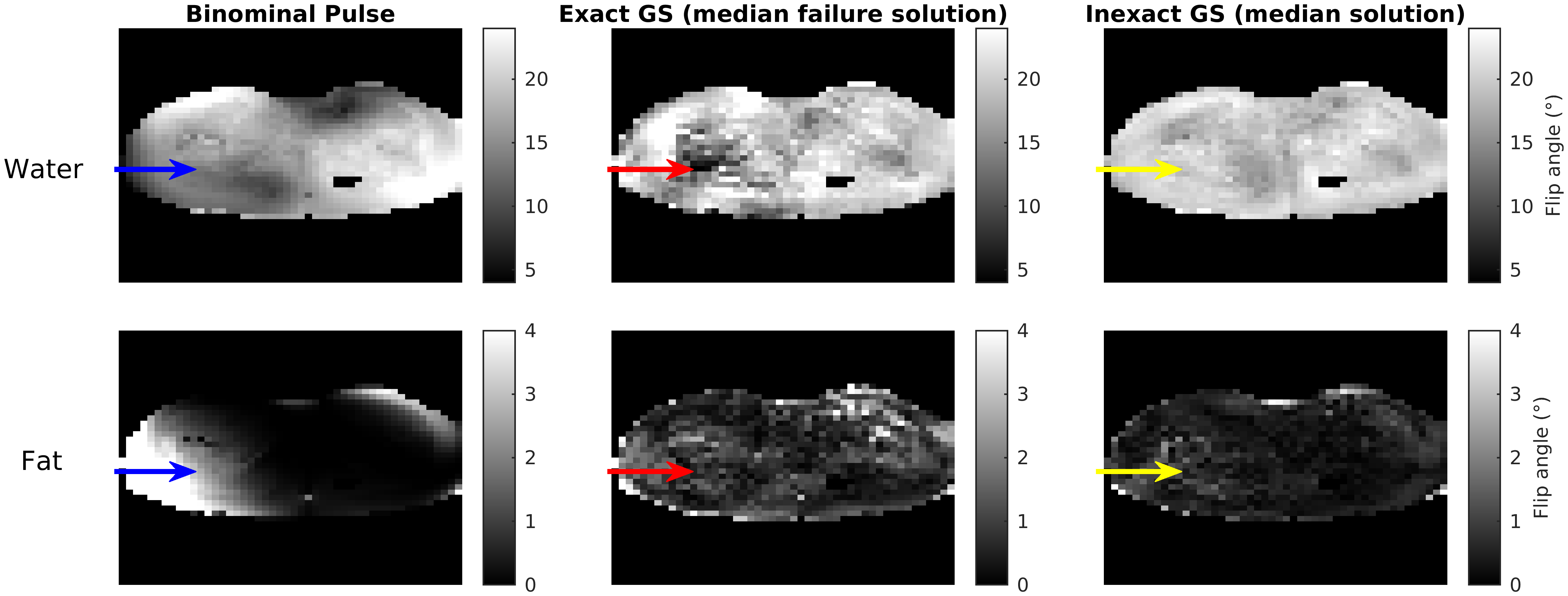}}
\caption{ { Simulation V: Spectral-Spatial 3T RF Pulse Design. Simulated water and fat excitations for a binominal 1-3-3-1 pulse and optimized pulses using exact GS and minibatched inexact GS. For exact GS, slice shown is  median failure slice (solutions with NRMSE $>0.18$), which had NRMSE = 0.1982. For inexact GS, slice shown is median-error slice (NRMSE = 0.0964). Colored arrows show the location where flip angles across frequency are plotted in Fig. \ref{fig_spec_spat_freq}}} 
\label{fig_spec_spat_slice}
\end{figure*}

\subsubsection{Simulation V: Spectral-Spatial 3T RF Pulse Design}
\begin{figure}[!t]
\centerline{\includegraphics[width=\columnwidth]{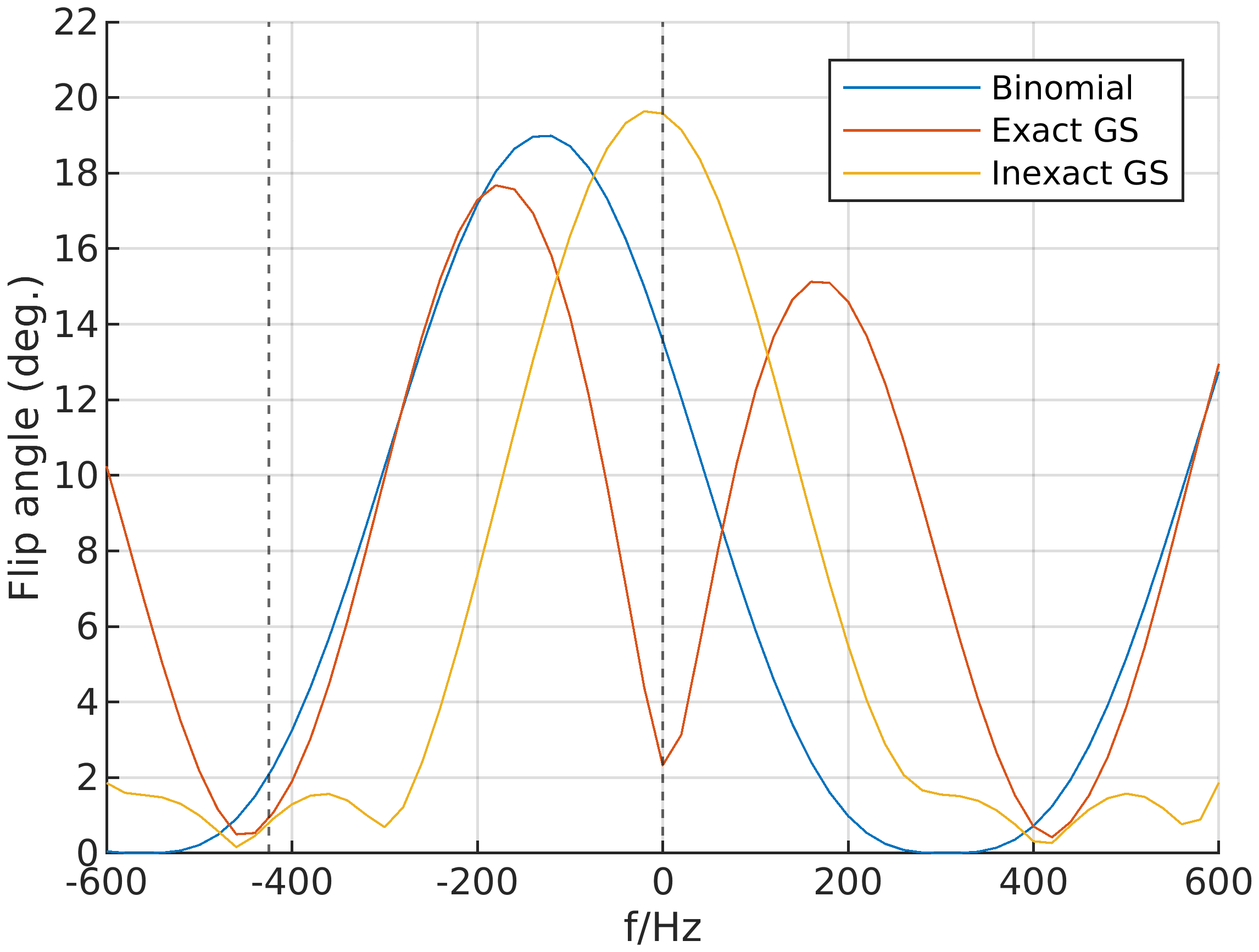}}
\caption{ { Simulation V: Spectral-Spatial 3T RF Pulse Design. Flip angle in degrees across the frequency spectrum, for spatial location denoted by arrows in Fig. \ref{fig4d15}. Dotted vertical lines show the locations of water (0 Hz) and fat (-425 Hz) frequencies at 3T. }}
\label{fig_spec_spat_freq}
\end{figure}

 { Figure \ref{fig_spec_spat_violin} shows the distribution of solution NRMSE values across 100 trials from the exact GS method and minibatched inexact GS method. Although inexact GS has a slightly higher median error solution than exact GS (exact = 0.0948, inexact = 0.0968), in 38\% of trials exact GS became stuck in a high-cost solution with NRMSE$>$0.18. These high-cost exact GS solutions are of poor quality and are considered failure trials. 
Figure \ref{fig_spec_spat_slice} shows the simulated water and fat excitations for the 1-3-3-1 binomial pulse and example results for the two optimized pulses. 
For the purposes of illustrating a failed solution, 
while the two methods had the same median error,
the exact GS solution that is shown is the median-cost failure trial; 
the inexact GS solution that is shown is the median-error overall. While both methods improved the excitation profile compared to the binomial pulse, 
the failure exact GS example shown has a large signal null at the water frequency and substantial out-of-band excitation at the fat frequency. 
In contrast, the median inexact GS solution does not have the same signal null in the water excitation and has less out-of-band excitation at the fat frequency. 
Signal nulls were not observed in any of the inexact GS results. 
The arrows in Fig. \ref{fig_spec_spat_slice} show the spatial location where the excitation across frequency is plotted in Fig. \ref{fig_spec_spat_freq}. 
The median inexact GS solution comes close to achieving the target water flip angle of 20$^\circ$, 
while the exact GS solution has a signal null centered at the water frequency with a flip angle of only 2.3$^\circ$.}

\subsection{Scanner Experiment}

Figure \ref{fig4d14} shows the slice-by-slice $B_1^+$  { coefficient of variation (CoV)} across the shimmed brain slices of the healthy volunteer. 
Both methods produced large improvements in $B_1^+$ homogeneity compared to their CP-mode initialization.
However in the middle slices of the volunteer's brain, 
using inexact GS produced a further reduction in $B_1^+$ CoV.

\par Figure \ref{fig4d15} shows predicted RF shim maps and measured GRE-EPI images for the slice indicated in Figure \ref{fig4d14}.  {The CP-mode shimmed slice has a conspicuous central brightening in the GRE-EPI image and reduced signal intensity in the back of the brain.}
There are several noticeable inhomogeneities in the exact-GS shimmed slice; 
conspicuous signal loss and signal hyperintensity are visible in two regions in the cortex, 
a region of interest in the GRE-EPI fMRI-type scan implemented. 
These signal inhomogeneities are reduced by inexact GS shimming and the overall slice $B_1^+$ CoV is reduced. 

\par
\begin{figure}[!t]

\centerline{\includegraphics[width=\columnwidth]{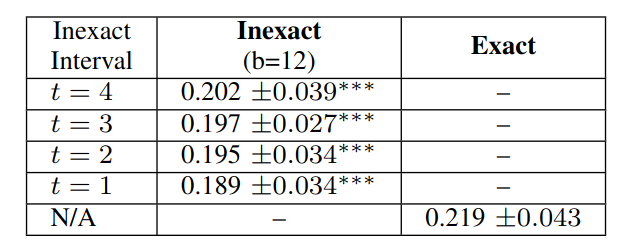}}
\caption{Table 2: Computational time per slice (s) for the in-vivo RF
shimming experiment.}
\label{tab2}
\end{figure}

  { Table \ref{tab2} shows the average per-slice design time in seconds for the scanner experiments using inexact and exact GS.
Using inexact GS with any of the examined inexact iteration intervals resulted in a highly significant reduction in computation time.}

\section{Discussion and Conclusion}
This study reported an approach to increase the robustness of magnitude-least squares RF pulse design and RF shimming. 
The method adapts the conventional Gerchberg-Saxton algorithm \cite{Setsompop2008MagnitudeChannels} by performing updates with a random sampling of pTx design system rows. 
This reduction of the pTx system makes solving the LLS subproblem  an inexact projection in the  Gerchberg-Saxton LLS problem resolution. 
As a result, these approximated projections can disrupt the convergence of Gerchberg-Saxton sequences, 
facilitating exploration of additional feasible points that could otherwise be missed. 
While the method itself does not converge independently, 
interleaving exact and inexact alternating projections balances local convergence and global exploration, 
yielding cost-effective solutions with significant image quality improvements.

\par
\begin{figure}[!t]
\centerline{\includegraphics[width=\columnwidth]{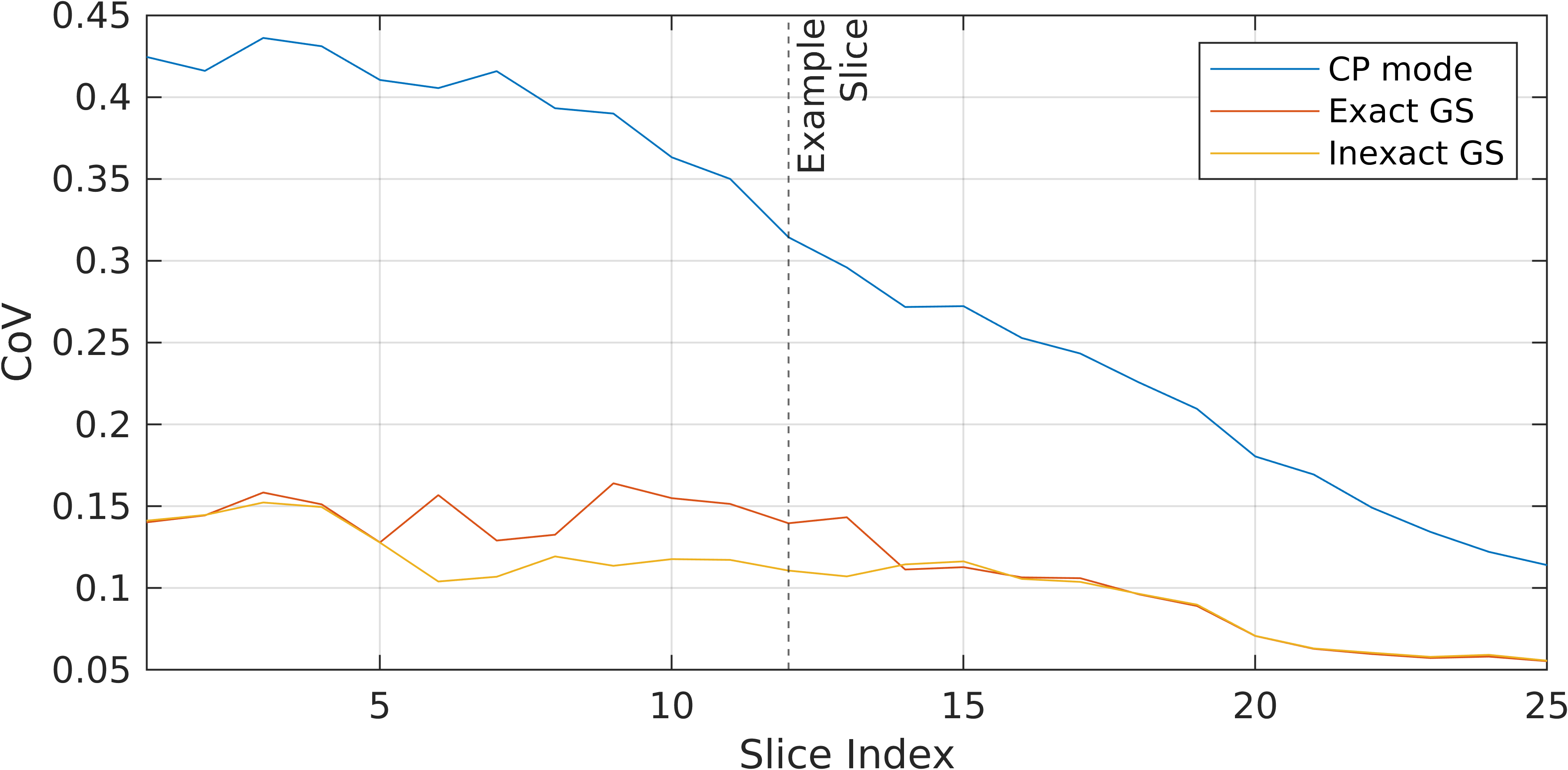}}
\caption{ {Scanner Results:} Slice-by-slice $B_1^+$ CoV from the in-vivo RF shimming experiment, for CP initialization, exact GS iterative optimization, and inexact GS optimization. Example slice in Fig. \ref{fig4d15} is marked. Inexact GS improved the CoV of the shims in the central slices of the head, 
with most improvement between slices 5 and 13.}
\label{fig4d14}
\end{figure}

\begin{figure*}[!t]
\centerline{\includegraphics[width=\textwidth]{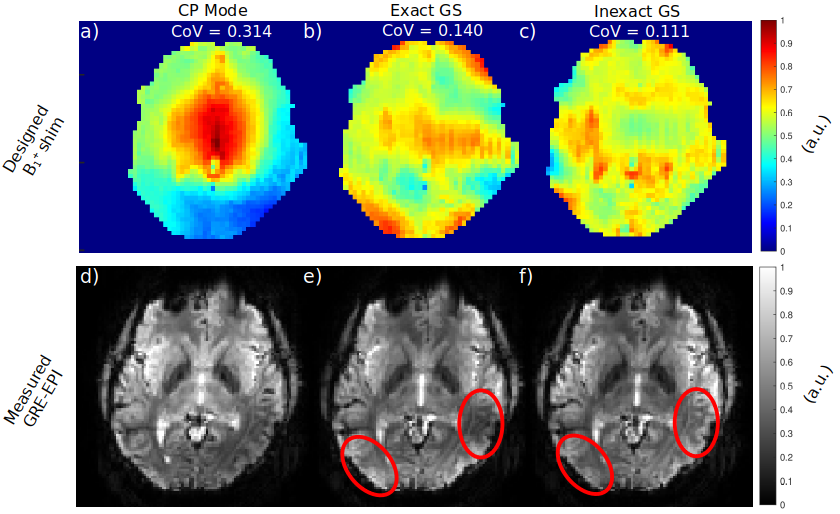}}
\caption{ {Scanner Results:} Compared  { designed $B_1^+$ RF shim} (a-c) and measured (d-f) signal in a shimmed slice acquired in the in-vivo experiment. The CP-mode shim has a central brightening with signal loss in the back of the head. 
Exact GS optimization reduces $B_1^+$ CoV but inhomogeneities remain (red). 
In the inexact GS solution, there is a further reduction in overall CoV and the signal reductions and hyperintensities noted in e) are resolved.}
\label{fig4d15}
\end{figure*}

 Compared to the conventional exact GS method, 
the inexact GS method consistently produced lower-cost solutions across sampled $B_1^+$ resolutions, transmit coil geometries,  { imaged anatomies, field strengths} and RF pulse types.  { Due to its' local convergence guarantees, 
conventional randomly-initialized exact GS may converge to any of the 
many local minima of the MLS optimization problem, some of which may be 
high-cost. 
In contrast, a minibatched inexact GS may escape local minima and find lower-cost solutions in general.
A consequence is that results across many trials with inexact GS may have lower variance than randomly-initialized exact GS, as shown in Figs. \ref{fig4} and \ref{fig5}}.
In the case of power-regularized optimization, these costs were lower in both RMSE and power cost, and fewer high-power outlier solutions were seen in Fig \ref{fig4d11}. 
 { Even in example cases in which the median exact GS solution was not significantly different than the median inexact GS solution, such as in Simulation V (the simulated spectral-spatial 3T RF pulse design), using inexact GS reliably prevented high-cost failure solutions as shown in Fig. \ref{fig4d14}.}
Although the stochastic nature of the method means that conventional well-initialized non-stochastic methods may occasionally outperform it, 
we found a high statistical likelihood that inexact GS produces superior solutions. 
 {For pulse design algorithms which are computationally intensive or for patient-tailored designs performed at the scanner, 
being able to avoid repeating trials to avoid high-cost solutions is desirable.}
Care must be taken to avoid numerical instability, 
but simple methods such as the addition of regularization or singular value thresholding can address these problems. 

\par Some of the more attractive features of the inexact GS approach are its simplicity and generalizability. 
There are very few parameters to tune, 
and those parameters that need to be determined (batchsize and inexact projection interval) have values that perform well over a wide range of datasets spanning a variety of $B_1^+$ resolutions and transmit arrays. 
This algorithm can be implemented by making only a few minor modifications to the already widely-used and extremely simple Gerchberg-Saxton pulse design method  { and was easily applied to the code base developed by an independent research group, Ref. \cite{MalikWater2010}}. 
In general, it is a favorable approach versus completely random perturbations since in the inexact iterations the search direction of the optimization in is still consistent with reducing cost over some subset of the spatial domain,  { an advantage in general of minibatch-based stochastic algorithms} \cite{Ruder2017AnAlgorithms}.  
Random initializations are not informed at all by the actual design problem, 
producing many trials that that may be initialized so far from low-cost solutions as to be useless. 
In contrast, the proposed method avoids the need to restart the optimization and a search that, while noisy, is always productive for some portion of the spatial domain. 
Inexact minibatched iterations are all data-dependent and at least somewhat consistent with minimizing over the entire set. 

\par We also observed that interleaving exact iterations that use all available $B_1^+$ data helped to further reduce cost versus performing only inexact updates. 
Interleaving exact iterations provides a means of regulating the variance of minibatch iterations in the absence of a mechanism to modify a gradient descent step size or otherwise temper the variance of minibatch gradient descent. 
Another possibility may simply be using larger batchsizes than those found to be optimal here with fewer or no interleaved exact iterations, which will help to reduce variance; the optimal choice balancing cost and computational expense will depend on the dimensionality of the problem.

\par We note that the emphasis in this work was on finding the solution with the lowest possible cost.
However, it is understood that the lowest cost RF pulse solution may not be a desirable one in some cases \cite{Paez2021RobustRegularization}. 
The simulated l-curve experiments shown in Figures \ref{fig4d10} and \ref{fig4d11} helped to demonstrate this; 
although the example excitation shown in Fig \ref{fig4d10} had lower overall cost than the solution in Fig \ref{fig4d11}, 
the solution with low power regularization had lower NRMSE and may be preferable if high power is permissible. 
Producing an excitation with the desired characteristics may require further modifications to the cost function
which are beyond the scope of the present work \cite{Paez2021RobustRegularization}.

\par One of the benefits of inexact GS is the potential for large reductions in computational cost. 
Since in the conventional exact GS approach the pseudoinverse of $\textbf{A}^H\textbf{A}$ is precomputed, 
the most significant computational burden is the vector-matrix multiplications required to solve the LLS optimization problem in each exact GS iteration, 
which takes at most $\mathcal{O}(mn)$ operations for a matrix $\textbf{A}\in\mathbb{C}^{m\times n}$ in naive implementations. 
Since the column dimension $n$ is usually small, 
shrinking $m$ can improve the speed of the algorithm, 
enabling more computationally intensive designs. 

\par In conclusion, a robust and efficient method was introduced for designing RF shims and pulses through magnitude-least-squares optimization. 
Simulations across diverse datasets and real-world scanning validation confirmed inexact projections' effectiveness. 
Future work will expand this MLS design algorithm to new pulse classes to assess other areas in which a minibatching approach can improve pulse design robustness and computational efficiency.

\appendices



\bibliography{main}

\end{document}